\documentstyle[12pt]{article}

\newcommand{\startappendix}{
\renewcommand{\thesection}{\Alph{section}}}

\let\sss=\l                     

\def\a{\alpha}
\def\b{\beta}
\def\c{\chi}
\def\d{\delta}
\def\e{\epsilon}                
\def\f{\phi}                    
\def\g{\gamma}
\def\h{\eta}

\def\j{\psi}
\def\k{\kappa}                  
\def\l{\lambda}

\def\o{\omega}
\def\p{\pi}                     
\def\q{\theta}                  
\def\t{\tau}

\def\F{\Phi}
\def\G{\Gamma}
\def\J{\Psi}
\def\L{\Lambda}

\def\P{\Pi}

\def\ca{{\cal A}}

\newcommand{\extraspace}{\addtolength{\abovedisplayskip}{2mm}
                        \addtolength{\belowdisplayskip}{2mm}
                        \addtolength{\abovedisplayshortskip}{2mm}
                        \addtolength{\belowdisplayshortskip}{2mm}}
\newcommand{\be}{\begin{equation}\extraspace}

\newcommand{\ee}{\end{equation}}

\newcommand{\bea}{\begin{eqnarray}\extraspace}
\newcommand{\beastar}{\begin{eqnarray*}\extraspace}
\newcommand{\eea}{\end{eqnarray}}
\newcommand{\eeastar}{\end{eqnarray*}}

\newcommand{\nonu}{\nonumber \\[2mm]}

\newcommand{\strutje}{\rule[-1.5mm]{0mm}{5mm}}

\newcommand{\dis}{\displaystyle}

\newcommand{\half}{\frac{1}{2}}

\newcommand{\wha}{\widehat{\cal A}}
\newcommand{\wta}{\widetilde{\cal A}}

\newcommand{\ind}{{\rm ind}}
\newcommand{\var}{\varphi}

\newcommand{\del}{\partial}

\newcommand{\bdel}{\bar{\partial}}

\newcommand{\osp}{OSp(N|2)}
\newcommand{\asp}{osp(N|2)}
\newcommand{\bz}{\bar{z}}
\newcommand{\slt}{sl(2,{\bf R})}
\newcommand{\sln}{sl(n,{\bf R})}
\newcommand{\unn}{\underline{n}}

\newcommand{\np}{Nucl.\ Phys.\ }

\newcommand{\cmp}{Comm.\ Math.\ Phys.\ }
\newcommand{\pl}{Phys.\ Lett.\ }
\newcommand{\pp}{=\!\!\! |}

\catcode`\@=11
\def\@afterindentfalse{\let\if@afterindent=\iftrue}
\catcode`\@=12

\begin{document}
\font  \biggbold=cmbx10 scaled\magstep2
\font  \bigbold=cmbx10 at 12.5pt
\font  \bigreg=cmr10 at 12pt

\bigreg

\baselineskip = 15pt

\noindent Mar. 1993 \hfill LBL-33738, UCB-PTH-93/06\\
$\strutje$ \hfill  KUL-TF-93/09\\
$\strutje$ \hfill  hep-th/9303133\\

\vspace{4mm}

\begin{center}
{\large INDUCED AND EFFECTIVE GRAVITY THEORIES IN D=2\footnote{This work was
supported in part by the Director,
Office of Energy Research, Office of High Energy and Nuclear Physics,
Division of High Energy Physics of the U.S. Department of Energy
under Contract DE-AC03-76SF00098 and in part by the National Science
Foundation under grant PHY90-21139.} }\\

\vspace{1cm}

{\bf \centerline{Alexander Sevrin${}^1$, Kris Thielemans${}^2$ and
Walter Troost\footnote{Bevoegdverklaard Navorser NFWO,Belgium}${}^{2}$}}
\vskip .3cm
{\baselineskip = 12pt
\centerline{\sl{1. Department of Physics}}
\centerline{\sl{University of California at Berkeley}}
\centerline{\sl{and}}
\centerline{\sl{Theoretical Physics Group}}
\centerline{\sl{Lawrence Berkeley Laboratory}}
\centerline{\sl{Berkeley, CA 94720, U.S.A.}}}
\vskip .2cm
{\baselineskip = 12pt
\centerline{\sl{2. Instituut voor Theoretische Fysica}}
\centerline{\sl{Universiteit Leuven}}
\centerline{\sl{Celestijnenlaan 200D, B-3001 Leuven, Belgium}}}
\end{center}

\vskip 1.cm
\centerline{\bf Abstract}
\vskip 0.5cm
{\baselineskip=14pt \small
As a preparation for the study of {\it arbitrary} extensions of $d=2$ gravity
we present a detailed investigation of $SO(N)$ supergravity. Induced $d=2$,
$SO(N)$ supergravity is constructed by gauging a chiral, nilpotent subgroup of
the $OSp(N|2)$ Wess-Zumino-Witten model. In order to get a gauge invariant
theory with the correct number of degrees of freedom, we need to introduce $N$
free fermions. From this we derive an all order expression for the effective
action. Reality of the coupling constant imposes the usual restrictions on $c$
for $N=0$ and 1. No such restrictions appear for  $N\geq 2$. For $N=2$, 3
and 4, no renormalizations of the coupling constant beyond one loop occur.
Also, the effective $N=4$ gravity based upon a
linear $N=4$
superconformal algebra, there is no renormalization at all, {\it i.e.} the
quantum theory is equal to the classical. These results are related to
non-renormalization theorems for theories with extended supersymmetries.
Arbitrary (super)extensions of $d=2$ gravity are then analyzed. The induced
theory is represented by a WZW model for which a chiral, solvable group is
gauged. From this, we obtain the effective action. All order expressions for
both the coupling constant renormalization and the wavefunction renormalization
are given. From this we classify all extensions of $d=2$ gravity for which the
coupling constant gets at most a one loop renormalization.
As an application of the general strategy, $N=4$ theories based on $D(2,1,\a)$
and $SU(1,1|2)$, all  $WA$ gravities and the $N=2$ $W_n$ models are treated in
some detail.}

\newpage

\renewcommand{\thepage}{\roman{page}}
\setcounter{page}{2}
\mbox{ }

\vskip 1in

\begin{center}
{\bf Disclaimer}
\end{center}

\vskip .2in

\begin{scriptsize}
\begin{quotation}
This document was prepared as an account of work sponsored by the United
States Government.  Neither the United States Government nor any agency
thereof, nor The Regents of the University of California, nor any of their
employees, makes any warranty, express or implied, or assumes any legal
liability or responsibility for the accuracy, completeness, or usefulness
of any information, apparatus, product, or process disclosed, or represents
that its use would not infringe privately owned rights.  Reference herein
to any specific commercial products process, or service by its trade name,
trademark, manufacturer, or otherwise, does not necessarily constitute or
imply its endorsement, recommendation, or favoring by the United States
Government or any agency thereof, or The Regents of the University of
California.  The views and opinions of authors expressed herein do not
necessarily state or reflect those of the United States Government or any
agency thereof of The Regents of the University of California and shall
not be used for advertising or product endorsement purposes.
\end{quotation}
\end{scriptsize}

\vskip 2in

\begin{center}
\begin{small}
{\it Lawrence Berkeley Laboratory is an equal opportunity employer.}
\end{small}
\end{center}

\newpage
\renewcommand{\thepage}{\arabic{page}}
\setcounter{page}{1}

\baselineskip=18pt

\section{Introduction}
\setcounter{equation}{0}
\setcounter{footnote}{0}

Soon after the observation of Polyakov \cite{polyakov,kpz} of a hidden, affine,
$Sl(2,{\bf R})$ symmetry in the effective $d=2$ gravity theory, $d=2$ gravity
was obtained from a gauged (or constrained) $Sl(2,{\bf R})$ Wess-Zumino-Witten
(WZW) model \cite{aleks,bo}. Following the understanding of $d=2$ gravity, a
major effort was put into the study of other types of $d=2$ gravity. Those
theories are based on extensions of the Virasoro algebra. Most of these
extensions are non-linearly generated algebras. These algebras have as
characteristic feature that the commutator of two generators contains not only
a linear combination of the generators but composites of the generators as
well:
\be
{[}T_a,T_b{]}=f_{ab}{}^cT_c+V_{ab}{}^{cd}T_cT_d+
              W_{ab}{}^{cde}T_cT_dT_e+\cdots \ .
\label{nola}
\ee
Though such algebras could be considered as ordinary Lie algebras by taking the
composites as new generators, such an approach is not very practical as it
often leads to an uncontrollable number of new generators.
Many properties of Lie algebras find their analogue in non-linearly generated
algebras \cite{ssvprep}. An important feature of these algebras is that a
distinction between classical and quantum algebras has to be made. Indeed, in
the classical case, the algebra is a Poisson bracket algebra and
composite terms in eq. (\ref{nola}) are unambiguously defined. In the quantum
case, the generators are operators on a Hilbert space and ordering ambiguities
arise. This fact is reflected in {\it e.g.} the Jacobi identities which
assume different forms in the classical and the quantum case \cite{brs2}.

A typical example of such algebras are $W$-algebras, which are higher spin
extensions of the Virasoro algebra. Since their discovery \cite{zamo},
$W$-algebras have attracted a lot of attention and numerous applications, both
in physics and mathematics, of this type of symmetry algebras have been found
(for a review see \cite{pbks}).

The prototype of these algebras is the $W_3$ algebra. It is generated by the
energy-momentum tensor $T$ and a dimension 3 current $W$ with operator product
expansions (OPE) given by:
\bea
T(x) T(y) &=& \frac{c}{2} (x-y)^{-4} + 2(x-y)^{-2} T(y)
    + (x-y)^{-1} \del T(y) + \cdots
\nonu
T(x) W(y) &=& 3 (x-y)^{-2} W(y) + (x-y)^{-1} \del W(y) +\cdots
\nonu
W(x) W(y) &=& \frac{c}{3} (x-y)^{-6}
    + 2 (x-y)^{-4} T(y) + (x-y)^{-3} \del T(y)
\nonu
&& + (x-y)^{-2} \left[ 2 \b \L (y) + \frac{3}{10} \del^2 T(y) \right]
\nonu
&& + (x-y)^{-1} \left[ \b \del \L (y) + \frac{1}{15} \del^3 T(y) \right]
   + \cdots ,
\label{w31}
\eea
where
\be
\L (x) = (TT) (x) - \frac{3}{10} \del^2 T(x)
\label{w32}
\ee
and
\be
\b = \frac{16}{22+5c}.
\label{w33}
\ee
Just as the Virasoro algebra appears as the residual symmetry of gauge fixed
gravity in $d=2$, the $W_3$-algebra appears as the residual symmetry of gauge
fixed $W_3$-gravity. The induced action for $W_3$-gravity in the chiral gauge
is
\be
e^{\dis -\G_\ind[h,b]} = \langle \exp - \frac{1}{\p} \int d^2 x \left[
h (x) T(x) + b(x) W(x) \right] \rangle \ .
\label{w3ind}
\ee
In \cite{ssvna}, it was shown that $\G_\ind[h,b]$ is expanded in
$1/c$ :
\be
\G_\ind[h,b]=\sum_{i\geq 0}c^{1-i}\G^{(i)}[h,b] \ .
\label{cexp}
\ee
This is in stark contrast with induced actions for linear conformal algebras,
which are proportional to $c$. The subleading terms in $1/c$ in eq.
(\ref{cexp}) arise from a proper treatment of the composite terms in eq.
(\ref{w31}). In \cite{ssvnb}, an explicit form for the classical term
$\G^{(0)}_\ind[h,b]$ was obtained through the classical reduction of an
$Sl(3, \bf{R})$ Wess-Zumino-Witten model
(WZW model).

The Legendre transform $W^{(0)}[t,w]$ of $\G^{(0)}_\ind[h,b]$ is defined by
\be
W^{(0)}[t,w]=\mbox{min}_{\{ h,b\}}\left( \G^{(0)}_\ind[h,b]-
\frac{1}{12\pi}\int d^2\, x\left( h\, t + \frac{1}{30} b\, w \right)\right).
\ee
In \cite{ssvnc}, it was conjectured that the generating functional $W[t,w]$ of
connected Green's functions,
defined by
\be
e^{\dis -W[t,w]} = \int [dh] [db] \,
  e^{\dis -\G_\ind[h,b]+\frac{1}{12\p} \int d^2 x\;\left(h t+\frac{1}{30} b
w\right) } .
\label{pi2}
\ee
is given by
\be
W[t,w] = \frac{k_c}{6}\, W^{(0)} \left[ Z^{(t)} t, Z^{(w)} w \right] ,
\label{exact}
\ee
where
\be
k_c = - \frac{1}{48} \left( 50-c-\sqrt{(c-2)(c-98)} \right) -3 \,,
\ee
and
\bea
Z^{(t)}=\frac{1}{2(k_c+3)} , \qquad
Z^{(w)}=\frac{1}{\sqrt{30\b}(k_c+3)^{3/2}} \ .
\eea
This conjecture was based on a computation of the first order quantum
corrections to $W^{(0)}$ which showed that the quantum corrections split into
two parts: one part contributes to the multiplicative renormalizations of
$W^{(0)}[t,w]$ while the other cancels $\G^{(1)}_\ind[h,b]$. If true, this
conjecture implies that the 1PI or effective action is simply given by
\be
\G_{\rm eff}[h,b] = \frac{k_c}{6} \, \G^{(0)} \left[ \frac{1}{2 k_c Z^{(t)}} h,
\frac{30}{2 k_c Z^{(w)}} b \right]  .
\ee
This result guarantees the integrability of $W_3$-gravity as the $Sl(3,{\bf
R})$ current algebra, essential to solve the theory, persists at quantum level.

Recently, this conjecture, for the case of $W_3$ at least, was elegantly proven
through the use of a quantum Hamiltonian reduction \cite{dbg}. The principle
behind this is quite simple and based on observations in \cite{aleks,bo}.
Consider a matter system, where we denote the matter fields collectively by
$\varphi$, with as action $S[\varphi ]$ and with a set of symmetry currents,
denoted by $J[\varphi ]$. The induced action is defined by
\be
e^{\dis -\G[A]}=\int[d\varphi]\mbox{e}^{\dis -S[ \varphi]-\frac 1 \p \int A
J[\varphi ]},
\ee
where $A$ is a source. Alternatively $A$ can be viewed as a chiral gauge
field. The generating functional of its connected Greens functions, which
upon a Legendre transform becomes the effective action, is defined by
\bea
e^{\dis -W[\tilde{J}]}&=&\int[d A]\mbox{e}^{\dis -\G[A]+\frac 1 \p \int A
\tilde{J}}\nonu
&=&\int [d\varphi]\d(J[\varphi ] -\tilde{J})\mbox{e}^{\dis -S[ \varphi]}.
\eea
Evaluating this functional integral is impossible in general as it
involves the
computation of a usually very complicated Jacobian. In the gravity case
however, one can realize the matter system by WZW model for which a chiral,
solvable group is gauged. In that case, the fields $\varphi$ are fixed
by the currents, and in addition the Jacobians are managable.

In this paper we will obtain an all order expression for the effective action
of an arbitrary extension of $d=2$ gravity. Before studying the general case,
we present a detailed study of $SO(N)$ supergravity. This case is most
instructive as it covers all subtleties encountered in the general case.

Aspects of $N=1$ and $N=2$ supergravity were studied in
\cite{grixu,sufrac,bo2}. The supergravity theories are based on the
$N$-extended $SO(N)$ superconformal algebra, which, for $N\geq 3$, is an
example
of a non-linearly generated algebra \cite{osp1,osp2}. The subalgebra of
transformations, globally defined on the sphere, form an $\osp$ algebra.
A realization of the matter sector referred to above is constructed from a
gauged $\osp$ WZW
model. Features of the $N=3$ theory were examined in \cite{gustav}, where
one-loop results for the effective action were given. We will give all loop
results for the effective theory for arbitrary $N$.

The paper is organized as follows. In the next section, the main properties of
induced $SO(N)$ supergravity are obtained through the study of the anomalous
Ward identities. In section 3, we get the large $c$ limit of the induced and
effective actions through the reduction of the $\osp$ WZW model.
In section 4, inspired by the methods of \cite{dbg}, an all order
representation of the induced action is constructed through a quantum
Hamiltonian reduction (or gauging) of the $\osp$ WZW model. In the following
section we use this representation to obtain an all order expression for the
effective action. The results are checked against one loop computations. In
section 6, we extend the framework to an arbitrary extension of $d=2$ gravity.
Both the coupling constant renormalization and the wavefunction renormalization
are explicitly computed. As an application we briefly analyze some other
$N=4$ supergravity theories. These are based on a one parameter family of
linear $N=4$ superconformal algebras \cite{n4}. By decoupling a $U(1)$ current
and 4 fermions \cite{goddard}, one obtains a one parameter family of
non-linearly generated $N=4$ superconformal algebras. The corresponding
supergravity theories are obtained by reducing the $D(2,1,\a)$ WZW model.  We
touch upon $SU(1,1|2)$ supergravity, all classes of $WA$ gravity and the $N=2$
extensions of $W_n$ gravity. We end by presenting some conclusions.
In appendix A,
we summarize several useful properties of WZW models on a supergroup. Induced
gauge theories are reviewed in appendix B. In appendix C we give a few useful
facts about $\slt$ embeddings.

\section{$SO(N)$ Supergravity}
\setcounter{equation}{0}
\setcounter{footnote}{0}

The $N$-extended $SO(N)$ superconformal algebras, \cite{osp1,osp2}, are
generated by the energy-momentum tensor $T$, $N$ dimension 3/2 supersymmetry
currents $G^a$ and an affine $SO(N)$ Lie algebra generated by currents $U^{i}$
where the index $i$ stands for a pair of indices $(pq)$ with $1\leq p < q \leq
N$. The operator product expansions (OPEs) are given by:
\bea
T(x) T(y) &=& \frac{c}{2} (x-y)^{-4} + 2(x-y)^{-2} T(y)
    + (x-y)^{-1} \del T(y)
\nonu
T(x) \F (y) &=& h_{\F} (x-y)^{-2} \F (y) + (x-y)^{-1} \del \F (y),
\nonu
G^a(x) G^b(y) &=& \d^{ab}b(x-y)^{-3}+2 \d^{ab} (x-y)^{-1} T(y)\nonu
&& + \frac{b}{k} \l_{ab}{}^i\left( 2(x-y)^{-2} U^{i}(y)+(x-y)^{-1}\del
U^{i}(y)\right)\nonu
&&+ (x-y)^{-1}\g\P_{ab}^{ij}(U^iU^j)(y),
\nonu
U^{i}(x)U^{j}(y)&=&-\frac k 2\d^{ij} (x-y)^{-2}+
(x-y)^{-1}f_{ij}{}^kU^k(y)\nonu
U^{i}(x)G^a(y)&=&(x-y)^{-1}\l_{ba}{}^iG^b(y),\label{alg2}
\eea
where
\bea
c&=&\frac k 2 \frac{6k + N^2-10}{k+N-3}\nonu
b&=&k\frac{2k+N-4}{k+N-3}\nonu
\g&=&\frac{2}{k+N-3}\label{relsc}
\eea
and
\be
h_{\F}=\frac 3 2 \mbox{ and } 1\mbox{ for }\quad \F=G^a \mbox{ and } U^{ab}.
\ee
The normalizations are such that $\l_{ab}{}^{(pq)}\equiv
1/\sqrt{2}(\d_a^p\d_b^q - \d_a^q\d_b^p)$, ${[}\l^i,\l^j{]}=f_{ij}{}^k\l^k$,
$tr(\l^i\l^j)=-\d^{ij}$, $f_{ik}{}^lf_{jl}{}^k=-(N-2)\d_{ij}$, and
$\P_{ab}^{ij}=\P_{ba}^{ij}=\P_{ab}^{ji} = \l_{ac}{}^{i}\l_{cb}{}^{j} +
\l_{ac}{}^{j}\l_{cb}{}^{i} + \d_{ab}\d^{ij} $. For $N=1$ and $N=2$ these are
just the standard $N=1$ and $N=2$ superconformal algebras. For $N\geq 3$ the
algebras contain composite terms in the $G\,G$ OPE.

The induced action $\G [h,\j,A]$ is defined as
\bea
\exp \left(-\G [h,\j,A]\right)&=&\Big\langle\exp\biggl( -\frac{1}{\pi} \int
d^2\,x \Bigl( h(x) T(x) + \j^a(x) G_a(x) \nonu
& & + A^{i}(x) U_{i}(x) \Bigr)\biggr)
\Big\rangle.\label{indn3}
\eea
The chiral, linearized supergravity transformations:
\bea
\d h&=&\bdel \e + \e \del h - \del \e h + 2 \q^a \j_a,
\nonu
\d \j^a&=&\bdel \q^a + \e \del \j^a - \frac 1 2 \del \e \j^a+\frac 1 2 \q^a\del
h-\del \q^a h\nonu
&& +\l_{ab}{}^i(\q^bA^i-\o^i\j^b),
\nonu
\d A^i&=&\bdel \o^i+\e \del A^i+\frac b k \l_{ab}{}^i(\del \q^a\j^b-\q^a\del
\j^b)
\nonu
&&-f_{jk}{}^i\o^jA^k-\del\o^ih,
\label{trans1}
\eea
are anomalous in the induced theory:
\be
\d  \G [h,\j,A ]=-\frac{c}{12\pi}\int\e\del^3h-\frac{b}{2\pi}\int\q^a\del^2\j_a
+ \frac{k}{2\pi}\int\o^i\del A_i-\frac{\g}{\pi} \int \q^a \j^b
\P_{ab}^{ij}\left(U^iU^j\right)_{\mbox{eff}}.\label{ano1}
\ee
The last term is due to the non-linear nature of the superconformal algebras.
Defining
\be
u^i=-\frac{2\pi}{k}\frac{\d\G [h,\j,A ]}{\d A^i}
\ee
one finds
\bea
\left(U^{(i}U^{j)}\right)_{\mbox{eff}}(x)&=&
\Big\langle\int d^2\,x\,U^{(i} U^{j)} (x)\exp\biggl( -\frac{1}{\pi} \int
\Bigl( h T + \j^a G_a + A^i U_a\Bigr)\biggr)
\Big\rangle/\exp \left(-\G\right),\nonu
&=&\left(\frac k 2 \right)^2 u^i(x)\,u^j(x)
+\frac{k\pi}{4}\lim_{y\rightarrow
x}\biggl(\frac{\del u^i (x)}{\del A^j(y)}
-\frac{\del}{\bdel}\d^{(2)}(x-y)\d^{ij} + i\rightleftharpoons j \biggr).
\label{effreg}
\eea
The limit in the last term of eq. (\ref{effreg}) reflects the point-splitting
regularization of the composite terms in eq. (\ref{alg2}). One notices that in
the limit $c\rightarrow\infty$, $u$ becomes $c$ independent and one has simply
\be
\lim_{c\rightarrow\infty}\left(\left( \frac 2 k \right)^2
\left(U^{(i}U^{j)}\right)_{\mbox{eff}}(x)\right)= u^i(x) \, u^j(x).
\label{effreglc}
\ee
Using eq. (\ref{effreg}), we find that eq. (\ref{ano1}) can be rewritten as:
\bea
\d  \G [h,\j,A ] &=& -\frac{c}{12\pi} \int\e\del^3h- \frac{b}{2\pi}\int
\q^a\del^2\j_a + \frac{k}{2\pi}\int\o^a\del A_a-\frac{k^2\g}{4\pi} \int \q^a
\j^b \P_{ab}^{ij} u^i u^j\nonu
&&-\lim_{y\rightarrow x}\frac{k\g}{2} \int \q^a \j^b \P_{ab}^{ij}
\biggl(\frac{\del u^i (x)}{\del A^j(y)}
-\frac{\del}{\bdel}\d^{(2)}(x-y)\d^{ij}\biggr).\label{ano3}
\eea
where the last term disappears in the large $k$ limit. The term proportional to
$\int\q^a\j^b\P^{ij}_{ab} u^i u^j$ in eq. (\ref{ano3}) can be absorbed by
adding a field dependent term in the transformation rule for $A$:
\be
\d_{\mbox{extra}}A^i=-\frac{k\g}{2}\q^{\,(a}\j^{b)}\P_{ab}^{ij}u_j.
\ee
Doing this, we find that in the large $k$ limit, the anomaly reduces to the
minimal one.

Introducing
\be
t=\frac{12\pi}{c}\frac{\d\G [h,\j,A,\h ]}{\d h}\qquad
g^a=\frac{2\pi}{b}\frac{\d\G [h,\j,A,\h ]}{\d \j^a}
\ee
we obtain the Ward identities by combining eqs. (\ref{trans1}) and
(\ref{ano1}):
\bea
\del^3h&=&\overline{\nabla}t-\frac{3b}{c}\left(\j^a\del + 3 \del\j^a\right)g_a
+\frac{6k}{c} \del A^iu_i,
\nonu
\del^2 \j^a&=&\overline{\nabla}g^a-\frac{c}{3b} \j^at+\l_{ab}{}^i A^ig^b
-\l_{ab}{}^i\left(2\del\j^b+\j^b\del\right)u^i
\nonu
&&-\frac{k^2\g}{2b}\P_{ab}^{ij}u^iu^j-
\frac{k\g\pi}{b} \P_{ab}^{ij}\lim_{y\rightarrow x}\biggl(\frac{\del u^i
(x)}{\del A^j(y)}
-\frac{\del}{\bdel}\d^{(2)}(x-y)\d^{ij} \biggr)\nonu
\del A^i&=&\overline{\nabla}u^i+\frac b k \l_{ab}{}^i\j^ag^b+f^{ij}{}_kA^ju^k,
\label{wi1}
\eea
where
\be
\overline{\nabla}\F=\left(\bdel-h\del-h_{\F}(\del h)\right)\F,
\ee
with
\be
h_{\F}=2,\frac 3 2 ,\ 1 \quad \mbox{ for }\quad \F=t,g^a,\ u^a.
\ee
The Ward identities provide us with a set of functional differential equations
for the induced action. Because of the explicit dependence on $k$ of the Ward
identities, the induced action is given as a $1/k$ expansion:
\be
\G [h,\j,A]=\sum_{i\geq 0}k^{1-i} \G^{(i)} [h,\j,A ].\label{ohwell}
\ee
In the large $k$ limit, the Ward identities become local and they are solved by
$\G [h,\j,A]=k\G^{(0)} [h,\j,A ]$. As we will show in the next section, the
large $k$ limit of the induced action can be obtained from a classical
reduction of an $\osp$ WZW model.

The effective action \footnote{For shortness, we will use the
term 'effective action' also for the generating functional of the connected
Green functions. We trust the symbol $W$ in stead of $\Gamma$ is enough to
avoid confusion.} $W[t,g,u]$ is defined by \bea
\exp \Big( -W[t,g,u]\Big) &=& \int [dh][d\j][dA]\exp\biggl(-\G [h,\j,A ]\nonu
&&+\frac{1}{4\pi}\Bigl(\int h\,t + 4 \j^a\,g_a -2 A^i\, u_i
\Bigr)\biggr).\label{qquan}
\eea
If one defines the Legendre transform of $\G^{(0)} [h,\j,A]$ as
\be
W^{(0)}[t,g,u]=\min_{\{h,\j,A\} } \left( \G^{(0)} [h,\j,A ]-
\frac{1}{4\pi}\left(\int h\,t + 4\j^a\,g_a -2  A^i\, u_i
\right)\right)\label{legdef}
\ee
we will show that $W[t,g,u]$ can be expressed as
\be
W[t,g,u]=k_cW^{(0)}[Z^{(t)}t,Z^{(g)}g,Z^{(u)}u].\label{renor}
\ee
So to leading ({\it i.e.} classical) order we have
\be
k_c\equiv k \qquad Z^{(t)}\equiv Z^{(g)}\equiv Z^{(u)}\equiv \frac 1 k
\ee


\section{Classical Reduction of $OSp(N|2)$}
\setcounter{equation}{0}
\setcounter{footnote}{0}

The Lie algebra of $\osp$ is generated by a set of bosonic generators
$\{t_{\pp}, t_0, t_=,t_{ab};t_{ab}=-t_{ba} \mbox{ and } a,b\in\{1,\cdots,N\}\}$
which form an $Sl(2)\times SO(N)$ Lie algebra and a set of fermionic generators
$\{t_{+a},t_{-a};a\in\{1,\cdots,N\}\}$. A Lie algebra valued field
$A=A^{\pp}\,t_{\pp}+A^{0}t_0+A^{=}t_= + \frac{1}{2} A^{ab}t_{ab} + A^{+a}t_{+a}
+ A^{-a}t_{-a}$ where the representation matrices are in the fundamental
representation, is explicitly given by:
\be
A \equiv \left( \begin{array}{ccccccc} A^0 & A^{\pp} & A^{+1} & A^{+2} & A^{+3}
& \cdots & A^{+N} \\ A^= & -A^0 & A^{-1} & A^{-2} & A^{-3} & \cdots & A^{-N} \\
A^{-1} & -A^{+1} & 0 & A^{12} & A^{13} & \cdots & A^{1N} \\ A^{-2} & -A^{+2} &
-A^{12} & 0 & A^{23} & \cdots & A^{2N}\\
A^{-3} & -A^{+3} & -A^{13} & -A^{23} & 0 & \cdots & A^{3N}\\
\vdots & \vdots & \vdots &\vdots & \vdots & & \vdots \\ A^{-N} & -A^{+N} &
-A^{1N} & -A^{2N} & -A^{3N} & \cdots & 0
\end{array} \right),
\label{ospmat}
\ee
and $A^{ab}=-A^{ba}$.
{}From this one reads off the generators of $\osp$ in the fundamental
representations and one easily computes the (anti)commutation relations.

Given the flat connections $A_{\bz}$ and $u_z$, {\it i.e.}
\be
R_{z\bz}=\del A_{\bz} -\bdel u_z-{[}u_z,A_{\bz}{]}=0,\label{curv}
\ee
we impose the following constraint on $u_z$
\be
u_z \equiv \left( \begin{array}{cccccc} 0 & u_z^{\pp} & u_z^{+1} & u_z^{+2} &
\cdots & u_z^{+N} \\ 1 & 0 & 0 & 0 & \cdots & 0 \\ 0 & -u_z^{+1} & 0 & u_z^{12}
&\cdots & u_z^{1N} \\ 0 & -u_z^{+2} & -u_z^{12} & 0 & \cdots & u_z^{2N}\\
\vdots & \vdots & \vdots & \vdots & & \vdots \\ 0 & -u_z^{+N} & -u_z^{1N} &
-u_z^{2N} &\cdots & 0
\end{array} \right).
\label{constr1}
\ee
After we impose the constraints eq. (\ref{constr1}), we find that some of the
components of $R_{z\bz}=0$ become algebraic equations. Indeed,
$R_{z\bz}^==R_{z\bz}^0=R_{z\bz}^{-a}=0$ for $0\leq a\leq N$ can be solved for
$A_{\bz}^0$, $A_{\bz}^{\pp}$ and $A_{\bz}^{+a}$ giving
\bea
A_{\bz}^0&=&\frac 1 2 \del A_{\bz}^=\nonu
A_{\bz}^{\pp}&=&-\frac 1 2 \del^2 A_{\bz}^=+A_{\bz}^=u_z^{\pp}
+ A_{\bz}^{-a}u_z^{+a}\nonu
A_{\bz}^{+a}&=&\del A_{\bz}^{-a}+A_{\bz}^{=}u_z^{+a} - \sqrt{2}\l_{ab}{}^i
A_{\bz}^{-b}u_z^i,
\eea
where
\be
u_z^i\equiv\frac{1}{\sqrt{2}}\l_{ab}{}^iu^{ab}_z.
\ee
The remaining curvature conditions $R_{z\bz}^{\pp} = R_{z\bz}^{+a} =
R_{z\bz}^{ab}=0$ reduce now to the Ward identities eq. (\ref{wi1}) in the limit
$k\rightarrow\infty$ upon identifying
\bea
h&\equiv&A_{\bz}^=\nonu
\j^a&\equiv&iA^{-a}_{\bz}\nonu
A^i&\equiv&-\sqrt{2}\left(A_{\bz}^i-A_{\bz}^=u^i_z\right)\nonu
t&\equiv&-2 \left(u_z^{\pp}+u^i_zu^i_z\right)\nonu
g^a&\equiv&iu^{+a}_z\nonu
u^i&\equiv&-\sqrt{2}u^i_z,
\eea
where
\be
A_{\bz}^i\equiv\frac{1}{\sqrt{2}}\l_{ab}{}^iA^{ab}_{\bz}.
\ee

Consider now a WZW model $\k S^-[g]$ (for conventions, see the appendix A) on
$\osp$. We can identify
\bea
&&W^{(0)}\left[  t= -2(\del g g^{-1})^{\pp} -2(\del g g^{-1})^i (\del g
g^{-1})^i, g^a= i (\del g g^{-1})^{+a} , u^i= - \sqrt{2}(\del g g^{-1})^{i}
\right]\nonu
&&\qquad\qquad\qquad\qquad\qquad\qquad =-\frac 1 2  S^-[g],\label{renor2}
\eea
where
$S^-[g]$ is the WZW-model with the constraints eq. (\ref{constr1}) imposed. As
such, one finds to leading order
\be
W[t,g,u]=kW^{(0)}[t/k,g/k,u/k]=\k S^-[g]
\ee
and the level $\k$ of the $\asp$ affine Lie algebra is related to the $SO(N)$
level $k$ as $k=- 2\k$ or $c=-6\k$.

Taking the Legendre transform
\be
\G [h,\j,A]=\min_{\{t,g,u\}}\left[
W [t,g,A ]+
\frac{1}{4\pi}\left(\int h\,t + 4\j^a\,g_a -2  A^i\, u_i  \right)
\right],
\ee
we find that induced action in the limit $k\rightarrow\infty$ is given by:
\be
\G [h,\j,A ]= k\G^{(0)} [h,\j,A ],\label{yy1}
\ee
where
\bea
&&\G^{(0)} \left[h=(\bdel g g^{-1})^=,\j^a=i(\bdel g g^{-1})^{-a},A^i=
-\sqrt{2}((\bdel g g^{-1})^i-(\bdel g g^{-1})^=(\del g g^{-1})^i_z)
\right]\nonu
&&= \frac 1 2 S^+[g]+\frac{1}{2\p}\int\left\{
(\bdel g g^{-1})^= \left( (\del g g^{-1})^{\pp}+(\del g g^{-1})^i (\del g
g^{-1})^i
\right)+
(\bdel g g^{-1})^{-a}(\del g g^{-1})^{+a}
\right\},\nonu
&&\label{yy2}
\eea
or
\be
\G [h,\j,A ]=-\k S^+[g]+\frac{\k}{2\p}\int\left(h t+ 2\j^ag^a\right).
\ee
In order to obtain an explicit expression for the induced action in the $k
\rightarrow \infty$ limit, we parametrize the group element $g$ as
\be
g\equiv \mbox{e}^{\dis x^{\pp}t_{\pp}}\, \mbox{e}^{\dis x^{+a}t_{+a}}
\,\mbox{e}^{\dis x^{0}t_{0}} \, \mbox{e}^{\dis \var^{i}t_{i}} \,\mbox{e}^{\dis
f t_{=}}\,\mbox{e}^{\dis \f^{b}t_{-b}},\label{yy3}
\ee
and we solve the constraints eq. (\ref{constr1}) with $u_z=\del g g^{-1}$.
This gives
\bea
x^0&=& \frac 1 2 \ln \left( \del f + \del\f^a \,\f^a\right)\nonu
x^{+a}&=&\left( \del f + \del\f^c \,\f^c\right)^{-1/2}\left[ \exp \left(
\sqrt{2} \var^i\l^i
\right)\right]_{ab}\del\f^b\nonu
x^{\pp}&=&-\frac 1 2 \del \ln \left( \del f + \del\f^a
\,\f^a\right).\label{yy4}
\eea
Using the parametrization eq. (\ref{yy3}) with the solutions, eq. (\ref{yy4}),
in eqs. (\ref{yy1}) (\ref{yy2}) one gets, through repeated application of the
Polyakov-Wiegman formula, eq. (\ref{pwfor}), the explicit form of the induced
action.

Finally, we can rewrite the right hand side of eq. (\ref{yy2}) in the following
suggestive form
\be
\G^{(0)} \left[h,\j,A\right]
= \frac 1 2 S^+[\mbox{e}^{\dis - z t_=} g]+\frac{1}{2\p}\int\left\{
(\bdel g g^{-1})^= (\del g g^{-1})^i (\del g g^{-1})^i\right\}
\label{yy8}
\ee
Defining
\be
\tilde{g}\equiv
\mbox{e}^{\dis - z t_=}g,
\ee
and using the Polyakov-Wiegman formula we get
\be
\G^{(0)} \left[h,\j,A\right]
= \frac 1 2 S^+[\tilde{g}]+\frac{1}{2\p}\int\left\{
h (\del \tilde{g} \tilde{g}^{-1})^i (\del \tilde{g} \tilde{g}^{-1})^i\right\},
\label{yy9}
\ee
where we have now that:
\bea
h&=&\left(\bdel \tilde{g}\tilde{g}^{-1}\right)^= + 2 z \left(\bdel
\tilde{g}\tilde{g}^{-1}\right)^0 -z^2 \left(\bdel
\tilde{g}\tilde{g}^{-1}\right)^{\pp}\nonu
\j^a&=&i\left(\bdel \tilde{g}\tilde{g}^{-1}\right)^{-a}+iz \left(\bdel
\tilde{g}\tilde{g}^{-1}\right)^{+a}\nonu
A^i&=&-\sqrt{2}\left(\bdel \tilde{g}\tilde{g}^{-1}\right)^i.
\label{h in J}
\eea
This clearly explains the origins of the hidden $\osp$ current algebra in
$SO(N)$ supergravity in the light-cone gauge.

This concludes our traatment of $SO(N)$ supergravity in the
$k\rightarrow\infty$ limit. In the next section, we will investigate this
theory for
arbitrary $k$.


\section{Quantum Reduction of $OSp(N|2)$}
\setcounter{equation}{0}
\setcounter{footnote}{0}

In order to obtain the full induced action, we consider a quantum Hamiltonian
reduction, {\it i.e.} instead of imposing the constraints eq. (\ref{constr1})
at the classical level we impose them as quantum conditions. Consider the
subalgebras $\P_\pm\asp$ of $\asp$ generated by
$\P_+\asp=\{t_{\pp},t_{+a};0\leq a\leq N\}$ and $\P_-\asp=\{t_{=},t_{-a}; 0\leq
a\leq N\}$. We call the corresponding subgroups of $\osp$, $\P_+\osp$ and
$\P_-\osp$. Introduce gauge fields $A_z\equiv \del h_- h_-^{-1}$ and
$A_{\bz}\equiv\bdel h_+ h_+^{-1}$ where $h_\pm\in \P_\pm\osp$. Obviously we
have that $A_z\in \P_-\asp$ and $A_{\bz}\in \P_+\asp$. The action
\be
{\cal S}_0=\k S^-[g]-\frac{\k}{2\p x} \int str A_zg^{-1}\bdel g +
\frac{\k}{2 \p x}\int str A_{\bz}\del g g^{-1} +\frac{ \k}{2 \p x}\int str
A_{\bz}g A_zg^{-1}\label{action1}
\ee
(with $x=\half$ in the fundamental representation) is invariant under
\bea
h_\pm &\rightarrow&\g_\pm h_\pm\nonu
g&\rightarrow&\g_+ g \g_-^{-1},\label{gauget1}
\eea
where
\be
\g_\pm\in\P_\pm\osp.
\ee
This can easily be shown by using the Polyakov-Wiegman formula, eq.
(\ref{pwfor}), to bring eq. (\ref{action1}) in the form ${\cal S}_0=\k
S^-[h_+^{-1}gh_-]$. In order to impose the constraints at the quantum
level, we would like to add the following term to the action
\be
+\frac \k \p \int  A_z^= - \frac \k \p \int  A_{\bz}^{\pp}.
\ee
However, as can easily be verified, this term is not invariant under the gauge
transformations eq. (\ref{gauget1}), which we take as our guiding
principle. From the form of the non-invariance terms,
one finds that invariance can be restored through the introduction of $2\times
N $ fermions, $\t^{+a}$ and $\t^{-a}$.  The action
\bea
{\cal S}_1 &=&{\cal S}_0-\frac \k \p\int \t^{+a}\bdel\t^{+a} -\frac \k \p\int
\t^{-a}\del\t^{-a}\nonu
&&-\frac\k\p\int\left(A^{\pp}_{\bz}+
2\t^{+a}A^{+a}_{\bz}\right)+\frac\k\p\int\left(A^{=}_{z}-
2\t^{-a}A^{-a}_{z}\right)
\eea
is fully invariant under eq. (\ref{gauget1}) provided the fermions transform as
\be
\t^{\pm a}\rightarrow\t^{\pm a}-\h^{\pm a},
\ee
and we parametrized
\be
\g_+\equiv\exp\left( \h^{\pp}t_{\pp}+\h^{+a}t_{+a}\right)\qquad
\g_-\equiv\exp\left( \h^{=}t_{=}+\h^{-a}t_{-a}\right)
\ee
The equations of motion for the fields $A^{\pp}_{\bz}$, $A^{+a}_{\bz}$,
$A^=_z$, $A^{-a}_{z}$ and $\t^{\pm a}$ read:
\bea
A^{\pp}_{\bz}&:&\left(\del g g^{-1}\right)^==1-\left( g A_z
g^{-1}\right)^=\nonu
A^{+a}_{\bz}&:&\left(\del g g^{-1}\right)^{-a}=\t^{+a}+\left( g A_z
g^{-1}\right)^{-a}\nonu
A^=_z&:&\left(g^{-1}\bdel g\right)^{\pp} = 1 +\left(g^{-1}
A_{\bz}g\right)^{\pp}\nonu
A^{-a}_{z}&:&\left(g^{-1}\bdel g\right)^{+a} = \t^{-a} +\left(g^{-1}
A_{\bz}g\right)^{+a}.
\label{eomg}
\eea
The two main options for fixing the gauge which are usually studied are the
conformal gauge and the light-cone gauge.
\begin{itemize}
\item Toda (or conformal) gauge

We use the gauge symmetry to restrict $g$ to ${\bf R}\times SO(N)$:
\be
g=\mbox{e}^{\dis \varphi t_0 + \f^{i}t_{i}},
\ee
The equations of motion for $A^{\pp}_{\bz}$, $A^{+a}_{\bz}$, $A^=_z$ and
$A^{-a}_{z}$, eq. (\ref{eomg}), become algebraic and we can perform the
integration over the $A_z$ and $A_{\bz}$ fields. The action becomes:
\bea
{\cal S}_1&=&-\frac \k \p \int\left( \del\var\bdel\var
+\t^{+a}\bdel\t^{+a}+ \t^{-a}\del\t^{-a}\right)-\k\widehat{S}_-[\hat{g}] \nonu
&&+\frac{\k}{\p}\int\left(\mbox{e}^{\dis 2\var}+2 \mbox{e}^{\dis \var} \t^{+a}
\hat{g}_{ab}\t^{-b}\right),
\eea
with
\be
\hat{g}_{ab}=\hat{g}^{-1}_{ba}\equiv\left[ \exp \left( \sqrt{2}\f^i\l^i
\right)\right]_{ab}
\ee
and where $\widehat{S}_-[\hat{g}]$ is an $SO(N)$ WZW model, {\it i.e.}
$\hat{g}=\exp\f^it_i=\exp\sqrt{2}\f^i\l_i$. The ``wrong'' sign in front of the
$\widehat{S}_-[\hat{g}]$
action is due to the fact that supertraces in the $SO(N)$ WZW action have been
replaced by ordinary traces.

This action describes the $N$-extended super Liouville theory or,
alternatively, the $\osp$ Toda action. A detailed study of super Liouville
theories will be presented elsewhere \cite{martin}, see also \cite{sorba}.

\item Drinfeld-Sokolov (or light-cone) gauge

We put
\be
A_z^==A_z^{-a}=0,\label{ppsym}
\ee
thus fixing the $\P_-\osp$ gauge symmetry and fix the $\P_+\osp$ gauge
symmetry by
\be
\left( \del g g^{-1}\right)^0=\t^{+a}=0.
\ee

{}From eq. (\ref{eomg}), one sees that the Lagrange multipliers $A^{\pp}_{\bz}$
and $A^{+a}_{\bz}$ impose the constraints
\be
(\del g g^{-1})^==1\qquad\qquad (\del g g^{-1})^{-a}=\t^{+a}. \label{hcon}
\ee
and we find that the constrained quantum current$\del g g^{-1}$ is of the
classical form given in eq. (\ref{constr1}). \end{itemize}

\vspace{.4cm}

We now claim that the system discussed above has an $N$-extended $SO(N)$
superconformal symmetry. In other words it possesses conserved currents
which
satisfy the OPEs given in eq. (\ref{alg2}). From now on we will essentially
work in the Drinfeld-Sokolov gauge or a slight modification thereof. We first
fix the $\P_-\osp$ gauge symmetry as in eq. (\ref{ppsym}) and modify the action
to:
\be
{\cal S}_1=\k S^-[g]-\frac \k \p\int \t^{+a}\bdel\t^{+a} + \frac \k \p \int
A_{\bz}^{\pp}\left(\left( \del g g^{-1} \right)^=-1\right)
-\frac{2\k}{\p}\int A^{+a}_{\bz} \left( \left(\del g
g^{-1}\right)^{-a}-\t^{+a}\right), \label{action2}
\ee
where we dropped the $-\frac \k \p\int \t^{-a}\del\t^{-a}$ term in the action
as it plays no  significant role in what follows.

The superconformal currents are those functionals which are invariant, modulo
the constraints eq. (\ref{hcon}), under the $\P_+\osp$ gauge transformations.
They are easily found by imposing the unique $\P_+\osp$ gauge transformations
which yields $(\del g g^{-1})^0 = \t^{+a}=0$. This transformation brings $(\del
g g^{-1})^{\pp}$, $(\del g g^{-1})^{+a}$, $(\del g g^{-1})^{i}$ in a form that
is proportional to the $SO(N)$ superconformal currents $T$ (modulo a term
proportional to the $SO(N)$ Sugawara tensor), $G^a$ and $U^i$. The explicit
form for the gauge invariant polynomials is
\bea
\tilde{J}^{\pp}_z&=&J^{\pp}+\frac 1 \a J^0J^0-2 J^{+a}\t^{+a}-
\sqrt{2}\l_{ab}{}^iJ^i\t^{+a}\t^{+b}
-\del J^0-\a\del \t^{+a} \t^{+a},\nonu
\tilde{J}^{+a}_z&=&J^{+a} -\sqrt{2}\l_{ab}{}^iJ^i
\t^{+b}-J^0\t^{+a}+\a\del\t^{+a}\nonu
\tilde{J}^{i}_z&=&J^i+\frac{\a}{\sqrt{2}}  \l_{ab}{}^i\t^{+a}\t^{+b},
\label{polys}
\eea
and we identified $J\equiv \a \del g g ^{-1}$ where classically $\a=\k /2$.
The polynomials given above are indeed invariant modulo the constraints:
\bea
\d \tilde{J}^{\pp}_z&=&-\frac\p 2\left(
\del\h^{\pp}+\h^{\pp}\del-\frac{4}{\k}\h^{\pp}J^0_{z}\right)\frac{\d{\cal
S}_1}{\d A^{\pp}_{\bz}}
\nonu
&&-\frac \p 4 \left(\h^{+a}\del + \del \h^{+a}+2\h^{\pp}\t^{+a}-2\h^{+b}
\t^{+b}\t^{+a}-\frac 4 \k \h^{+a}J^{0}_z\right)\frac{\d{\cal S}_1}{\d
A^{+a}_{\bz}}\nonu
\d\tilde{J}^{+a}_z&=&-\frac \p 2\h^{\pp}\t^{+a} \frac{\d{\cal S}_1}{\d
A^{\pp}_{\bz}}-\frac \p 4\left(
\h^{\pp}\d_{ab}+\h^{+a}\t^{+b}-\h^{+b}\t^{+a}-\d_{ab}\h^{+c}\t^{+c}
\right)\frac{\d{\cal S}_1}{\d A^{+b}_{\bz}}
\nonu
\d \tilde{J}^{i}_z& =&-\frac{\p}{2\sqrt{2}}\l_{ab}{}^i\h^{+a}\frac{\d{\cal
S}_1}{\d A^{+b}_{\bz}}.\label{noninvar}
\eea

Having found the currents realising $SO(N)$ supergravity, our aim is now to
obtain the induced and effective action through this realisation, following
the method used for $W_3$ in \cite{dbg}.
To this end we couple the currents to sources $h$, $\j^a$ and $A^i$,
and consider the classical action ${\cal S}_2$:
\be
{\cal S}_2={\cal S}_1+\frac 1 \p\int\left( h T+\j^aG_a+ A^iU_i \right)
\ee
with ${\cal S}_1$ was defined in eq. (\ref{action2}) and $T$, $G^a$ and $U^i$,
\bea
T&=&C_T\left(\tilde{J}^{\pp}_z+\frac 2 \k
\tilde{J}^{i}_z\tilde{J}^{i}_z\right)\nonu
G^a&=&C_G \tilde{J}^{+a}_z\nonu
U^i&=&C_U\tilde{J}^{i}_z,\label{current1}
\eea
where $C_T=2$, $C_G=4i$ and $C_U=-2\sqrt{2}$, satisfy the classical limit of
the superconformal algebra eq. (\ref{alg2}).

{}From eq. (\ref{noninvar}) we get that the couplings to the sources $h$,
$\j^a$ and $A^i$ in the action are gauge invariant up to terms proportional to
the equations of motion of $A^{\pp}_{\bz}$ and $A^{+a}_{\bz}$. We can cancel
these noninvariance terms by adding extra terms to the transformation rules of
$A^{\pp}_{\bz}$ and $A^{+a}_{\bz}$. As $A^{\pp}_{\bz}$ and $A^{+a}_{\bz}$
appear linearly in the action, no further modifications are needed and the
action is invariant under $\P_+\osp$ gauge transformations. In this way we
obtain a realization of the induced action:
\be
\exp -\G[h,\j,A]=\int [\d g g^{-1}][d\t][d A_{\bz}]\left( \mbox{Vol}\left(
\P_+\osp \right) \right)^{-1}\exp-{\cal S}_2[g,\t,A_{\bz}],\label{okok}
\ee
provided that at quantum level the currents eq. (\ref{current1}), up to
multiplicative renormalizations and terms containing ghostfields, satisfy the
quantum $SO(N)$ superconformal algebra.

The gauge fixing procedure is most easily performed using the
Batalin-Vilkovisky (BV) method \cite{bv}. We skip the details here, a readable
account of the BV method can be found in {\it e.g.} \cite{proeyen}. For each of
the fields appearing in the theory we introduce anti-fields of opposite
statistics which we denote by $J_z^*$, $A^*_{\bz}$, $c^{\pp\, *}$, etc. The
solution to the master equation is given by:
\bea
{\cal S}_{\rm BV}&=&{\cal S}_{2}-\int J_z^{-a*}\g^{+a}J^=_z+ \int
J_z^{0*}\left( c^{\pp}J_z^=+\g^{+a}J_z^{-a}
\right)-\sqrt{2}\l_{ab}{}^i\int J_z^{i*}\g^{+a}J_z^{-b}\nonu
&&-\int J_z^{+a*}\left(\g^{+a}J_z^0+c^{\pp}J_z^{-a} +\sqrt{2}
\l_{ab}{}^i\g^{+b}J^i_z-\frac\k 2 \del\g^{+a}
\right)\nonu
&&+\int J_z^{\pp \, *}\left(\frac\k 2 \del c^{\pp}-2 c^{\pp}J_z^0 - 2\g^{+a}
J_z^{+a}\right)-\int \t^{+a*}\g^{+a}+\int c^{\pp\, *}\g^{+a}\g^{+a}\nonu
&&+\int A_{\bz}^{\pp\, *}\left( \bdel c^{\pp}-2\g^{+a}A_{\bz}^{+a}-c^{\pp}\del
h -\frac 4 \k c^{\pp}h J_z^0+2i c^{\pp}\j^a\t^{+a}
\right)\nonu
&&+\int A_{\bz}^{+b *}\left(\bdel\g^{+b}-\frac 1 2 \g^{+b}\del h
-c^{\pp}h\t^{+b} -\frac 2 \k\g^{+b}h J_z^0
+ \frac{2\sqrt{2}}{\k}\l_{ab}{}^i\g^{+a} J_z^ih\right.\nonu
&&- ic^{\pp} \j^b -i\g^{+a} \j^a \t^{+b}+i\g^{+b} \j^a \t^{+a}+i\g^{+a} \j^b
\t^{+a}
-\l_{ab}{}^i \g^{+a} A^i\Biggr)\nonu
&&+ \frac\p\k\int A^{\pp\,*}_{\bz}
A^{+a*}_{\bz}\g^{+a}c^{\pp}h+\frac{\p}{4\k}\int A^{+a*}_{\bz}A^{+a*}_{\bz}
\g^{+b}\g^{+b}h.\label{bvresult}
\eea

In order to compute the normalization of the currents in eq. (\ref{current1})
and $k$ as a function of $\k$ in the quantum theory, we choose a hybrid gauge
to fix the $\P_+\osp$ gauge invariance by putting $A^{\pp}_{\bz} =
A^{+a}_{\bz}=0$, which is different from the Drinfeld-Sokolov gauge! This gauge
choice will allow us to compute the normalization constants and $k(\k )$ using
operator methods. The all-order computation of these constants, using solely
functional methods, seems technically not feasible.

To fix the gauge, we make a canonical transformation (of fields and
antifields) such that the functional integration over the {\it new} fields
has no gauge-directions. Here this transformation is very simple, it
consists in interchanging the denominations 'field' and 'antifield' for the
canonical pairs $\{A^{\pp}_{\bz},A^{\pp\,*}_{\bz}\}$ and
$\{A^{+a}_{\bz},A^{+a*}_{\bz}\}$. Introducing more conventional names
for the Faddeev-Popov antighosts (which is what they turn out to be, in
this gauge), we put
\bea
   A^{+a*}_{\bz   } &=&-\frac{2}{\p}\b^{-a}     \nonu
   A^{+a}_{\bz}     &=&\frac{\p}{2}\b^{-a*}     \nonu
   A^{\pp\,*}_{\bz} &=&\frac{b^=}{\p}           \nonu
   A^{\pp}_{\bz}    &=&-\p b^{=*}
\eea
The gauge fixed action is then simply obtained from ${\cal S}_{\rm BV}$ by
putting the (new) antifields to zero. One finds:
\be
{\cal S}_{\rm gf}=\k S^-[g]-\frac \k \p\int
\t^{+a}\bdel\t^{+a}+\frac{1}{\p}\int b^=\bdel c^{\pp}-\frac 2 \p\int
\b^{-a}\bdel \g^{+a}+\frac 1 \p\int\left( h \tilde{T}+\j^a \tilde{G}_a+
A^i\tilde{U}_i \right),\label{gauges1}
\ee
where $\tilde{T}$, $\tilde{G}^a$ and $\tilde{U}^i$ have precisely the form
given in eq. (\ref{current1}) but with the currents $J^{\pp}_z$, $J^{+a}_z$,
$J^0_z$ and $J^{i}_z$ replaced by $\hat{J}^{\pp}_z$, $\hat{J}^{+a}_z$,
$\hat{J}^{0}_z$ and $\hat{J}^{i}_z$, given by
\bea
\hat{J}^{\pp}_z&=&J^{\pp}_z\nonu
\hat{J}^{+a}_z&=&J^{+a}_z-\frac 1 2 c^{\pp}\b^{-a}\nonu
\hat{J}^{0}_z&=&J^{0}_z-\frac 1 2 b^= c^{\pp}+\frac 1 2 \b^{-a}\g^{+a}\nonu
\hat{J}^{i}_z&=&J^{i}_z+ \frac{1}{\sqrt{2}} \l_{ab}{}^i \b^{-a}
\g^{+b}.\label{current3}
\eea
Note that the new extended action is linear in antifields, which makes the
gauge
algebra in fact simpler than in the original variables, closing now even
off shell. Only a few terms that will be needed later are repeated here
explicitly:
\bea
{\cal S}_{\rm BV}&=&{\cal S}_{\rm gf}
       -\k b^{=*}\left((\del g g^{-1})^{=}-1\right)
       -\b^{-a*} \left(\k(\del g g^{-1})^{-a}-\k\t^{+a}
       +b^{=}\g^{+a}\right) +\cdots \label{extactgauge1}
\eea

This ends the construction of the classical extended and gauge fixed
action. To preserve the gauge invariance at the quantum level, it may be
necessary to add quantum corrections to this extended action. We will not
make a fully regularised quantum field theory computation.
To make the transition to the quantum theory, we use BRST invariance as a
guide. We will use OPE-techniques without specifying a regularisation
underlying this method in renormalised perturbation theory.

By construction, the action, eq. (\ref{gauges1}) is classically BRST
invariant with the BRST charge given by
\be
Q=\frac{1}{2\p i}\oint\left\{c^{\pp}\left(J_z^=-\frac{\k}{2}\right)+ 2\g^{+a}
\left(
J_z^{-a}-\frac \k 2 \t^{+a}\right) -\frac 1 2 b^= \g^{+a}\g^{+a}.
\right\}
\ee
Using the $\osp$ OPEs given in eq. (\ref{ospope}) and
\bea
\t^{+a}(x)\t^{+b}(y)&=&-\frac{1}{2\k}\d^{ab}(x-y)^{-1}\nonu
b^=(x) c^{\pp}(y)&=&(x-y)^{-1}\nonu
\b^{-a}(x) \g^{+b}(y) &=&\frac 1 2 \d^{ab} (x-y)^{-1},
\eea
one finds that the BRST charge is also in the  quantum theory nilpotent,
provided that the currents still satisfy the OPE's of eq. (\ref{ospope}).
It is known that the classical relation
$J=\frac{\k}{2}\del g g^{-1}$ is then renormalised to
$J=\frac{\a_\k }{2}\del g g^{-1}$, where the analysis of WZW models in
the operator formalism \cite{kz}, strongly suggests that $\a_\k=\k+\tilde{h}$.
{}From now on we do choose the value $\a_\k=\k+\tilde{h}$.
It may also be noted that the hatted currents
satisfy the same algebra as the unhatted ones in eq.
(\ref{ospope}), with only the following modifications in the
central terms:
\bea
\hat{J}^0(x)\hat{J}^0(y)&=&\frac{ 2\k+4-N}{16} (x-y)^{-2}\nonu
\hat{J}^{i}(x)\hat{J}^{j}(y)&=&\frac {\k+1}{ 8} \d^{ij}(x-y)^{-2}-\frac{
\sqrt{2}}{4}(x-y)^{-1} f_{ij}{}^k\hat{J}^k(y).
\label{hospope}
\eea

Now we construct the quantum corrections to the action by using the BRST
invariance. The gauge fixed action for $h=\j^a=A^i=0$ is invariant as it
stands. Since $h$, $\j^a$ and $A^i$ do not
transform, we determine the quantum form of the $T$, $G^a$ and $U^i$
currents by requiring them to be BRST invariant also.
This results in the following currents:

\bea
\tilde{T}&=&C_T\Biggl(\hat{J}^{\pp}_z+2\t^{+a}\hat{J}^{+a}_z +\frac 2 \k
\hat{J}^0_z\hat{J}^0_z -\frac{\k+1}{\k}\del \hat{J}^0_z +
\frac{2}{\k}\hat{J}^i_z \hat{J}^i_z  \Biggr)- \k\del \t^{+a} \t^{+a}
\nonu
\tilde{G}^a&=&C_G\left( \hat{J}^{+a}_z -\sqrt{2}\l_{ab}{}^i\t^{+b}\hat{J}^i_z
-\t^{+a}\hat{J}^0_z+\frac{\k+1}{2}\del\t^{+a}\right)\nonu
\tilde{U}^i&=& C_U\left(\hat{J}^i_z+\frac{\k}{2\sqrt{2}}
\l_{ab}{}^i\t^{+a}\t^{+b}\right),\label{current2}
\eea
where
\bea
C_T&=&\frac{4\k}{2\k+4-N}\nonu
C_G&=&\sqrt{\frac{32\k}{N-2\k-4}}\nonu
C_U&=&-\frac{4}{\sqrt{2}}, \label{Cfactors}
\eea
and the bilinears are understood as regular parts in the OPE-expansions.

These currents satisfy the superconformal algebra eq. (\ref{alg2}) with
\be
k=-2\k -1.
\ee
Therefore, we showed that the reduced $\osp$ WZW model,
eq. (\ref{okok}), yields a representation of induced $SO(N)$ supergravity
at the quantum level.

To close, we also give the quantum corrections in the extended action
to some other terms that are needed in the next section. In the present
gauge, these correspond to terms proportional to
antifields, i.e. to transformation laws. Since we know the BRST charge
explicitly, the quantum transformation laws are easy to derive in this
gauge.
This entails the following modifications of eq.(\ref{extactgauge1}):
\bea
{\cal S}_{\rm BV}^{q}
    &=&{\cal S}_{\rm gf}^{q}
       -b^{=*}
           \left(\a_\k (\del g g^{-1})^{=}-\k\right)
        -\b^{-a*} \left(\a_\k (\del g g^{-1})^{-a}-
          \k\t^{+a}+b^{=}\g^{+a}\right) + \cdots     \nonu
    &=&{\cal S}_{\rm gf}^{q}
       +\frac{A^{\pp}_{\bz}}{\p}
           \left(\a_\k(\del g g^{-1})^{=}-\k\right)
        -\frac{2A^{+a}_{\bz}}{\p}
        \left(\a_\k(\del g g^{-1})^{-a}-
          \k\t^{+a}+b^{=}\g^{+a}\right) + \cdots     \nonu
          \label{Qextactgauge1}
\eea
where ${\cal S}^{q}_{\rm gf}$ refers to the gauge fixed action, with
the modified currents of eq.(\ref{current2}), and we reverted to the
original names of the ghost-variables.

{}From the BV viewpoint, the modification of the currents,
eq.(\ref{current2}) and the transformation laws, eq.(\ref{Qextactgauge1})
amount to (part of) the computation of the quantum corrections ($M_i$) to
the classical extended action. Without computing $\Delta S$ and checking
the quantum BV master equation, which would require some regularisation
procedure, we feel confident that $Q^2=0$ guarantees the gauge invariance
of the quantum theory. Therefore we will use
eq.(\ref{Qextactgauge1}) as it stands also for a different gauge.

\section{The Effective Action}
\setcounter{equation}{0}
\setcounter{footnote}{0}

\subsection{All Order Results}

The expression for the induced action which we obtained in previous section, is
perfectly suited for an all order computation of the effective action. The
effective action was defined in eq. (\ref{qquan}) where $\G [h,\j,A ]$ is given
in eq. (\ref{okok}). In order to compute it, we fix the $\P_+\osp$ gauge
invariance by choosing the Drinfeld-Sokolov gauge, {\it i.e.}
$J_z^0=\t^{+a}=0$. In the BV formalism, this is achieved by
turning them into antifields. So we again make a simple canonical
transformation interchanging fields and antifields, this time on the
canonical pairs $\{J^0_z,J^{0*}_z\}$ and $\{\t^{+a},\t^{+a*}\}$, i.e. we
now put
\bea
   J^{0*}  &=&\frac{b^=}{\p}            \nonu
   J^0     &=&-\p b^{=*}                \nonu
   \t^{+a*}&=&-\frac{2}{\p}\b^{-a}      \nonu
   \t^{+a} &=&\frac{\p}{2}\b^{-a*}      \nonu
\eea
Note that some of the quantum corrections to the transformation laws,
eq.(\ref{Qextactgauge1}), {\it e.g.} the fact
that $J_z=(\k+\tilde{h})/2\del g g^{-1}$ and the normalization of the
leading
terms of the superconformal currents eq. (\ref{Cfactors}), now show up in
the gauge fixed action itself.

Combining eqs. (\ref{qquan}), (\ref{okok}) and (\ref{bvresult}), the action
becomes:
\bea
&&\exp -W[t,g,u]=\int [d(\d gg^{-1})][d\t][dA_{\bz}][db][dc][d\b][d\g][dh][d\j
][dA]
\d\left(\t^{+a}\right) \d\left(J_z^0\right) \nonu
&&\qquad \exp\Bigg(-\k S^-[g]+\frac \k \p \t^{+a}\bdel\t^{+a}
- \frac 2 \p \int  A_{\bz}^{\pp}\left( J^=_z-\frac \k 2 \right)
+\frac{4}{\p}\int A^{+a}_{\bz} \left( J^{-a}_z-\frac \k 2 \t^{+a}\right)
\nonu
&&\qquad -\frac 1 \p \int b^=\left( c^{\pp}J^=_z+\g^{+a}J^{-a}_z\right)-\frac 2
\p \int \b^{-a}\g^{+a}\nonu
&&\qquad-\frac 1 \p \bigg(\int h \left( T-\frac t 4 \right)\nonu
&&\qquad+\j^a \left( G_a-g_a\right)+
 A^i\left(U_i+\frac 1 2 u_i\right) \bigg) \Bigg)
\eea
Passing from the Haar measure $[\d g g^{-1}]$ to the measure $[dJ_z]$, see eq.
(\ref{yoyo}), we pick up a Jacobian:
\be
[\d g g^{-1}]=[dJ_z]\exp \left( (N-4) S^-[g]\right).\label{chmeas}
\ee
Performing the integration over the $A_{\bz}$, $\t$, $h$, $\j$ and $A$, we get:
\bea
&&\exp -W[t,g,u]=\int[dJ_z][db][dc][d\b][d\g]\d\left(J_z^=-\frac \k 2 \right)
\d\left(J_z^{-a}\right) \d\left(J_z^{0}\right)\nonu
&&\qquad \d\left( C_T J^{\pp}_z+\frac{ 2C_T}{\k } J^i_zJ^i_z - \frac t
4\right)\d\left( C_GJ_z^{+a}-g^a\right) \d\left( C_UJ^i_z+\frac{u^i}{2}\right)
\exp\bigg(-\k_cS^-[g]\nonu
&&\qquad -\frac 1 \p \int b^=\left( c^{\pp}J^=_z+\g^{+a}J^{-a}_z\right)-\frac 2
\p \int \b^{-a}\g^{+a}\bigg)\label{oo1}
\eea
where
\be
\k_c=\k+4-N.\label{newc}
\ee
Performing the integrals over $J_z$ and then over the ghosts in eq.
(\ref{oo1}), we observe that the ghost contributions amount to an overall
factor, which we drop. Then we get that $W[t,g,u]$ is given by
\be
W[t,g,u]=\k_c S^-[g],
\ee
where the WZW functional is constrained by:
\bea
\left(\del g g^{-1}\right)^=&=&\frac{\k}{\a_\k}\nonu
\left(\del g g^{-1}\right)^{-a}&=&\left(\del g g^{-1}\right)^0=0\nonu
\left(\del g g^{-1}\right)^{\pp}+\frac{\a_\k}{\k}\left(\del g
g^{-1}\right)^i\left(\del g g^{-1}\right)^i&=&\frac{t}{2\a_\k C_T}\nonu
\left(\del g g^{-1}\right)^{-a}&=&\frac{2}{\a_\k C_G}g^a\nonu
\left(\del g g^{-1}\right)^{i}&=&-\frac{1}{\a_\k C_U}u^i\label{qpq}
\eea
Comparing with the results of section 3, we find that we rather need the
constraint $(\del g g^{-1})^==1$ instead of $(\del g g^{-1})^==\k/\a_\k$.
The conversion is easily performed through a global group transformation:
\be
g\rightarrow \mbox{e}^{\dis \ln \left(\sqrt{\frac{\a_\k}{\k}}\,\right) t_0 }g.
\ee
Combining this with eqs. (\ref{renor}) and (\ref{renor2}), we get
\be
W[t,g,u]=-2\k_c W^{(0)}\left( Z^{(t)}t,Z^{(g)}g,Z^{(u)}u \right),
\ee
where
\be
\k_c=-\frac 1 2 \left( k-7+2N\right), \label{relsd}
\ee
and
\bea
Z^{(t)}&=&-\frac{\k}{C_T\a_\k^2}\nonu
Z^{(g)}&=&\frac{2i}{C_G\a_\k}\sqrt{\frac{\k}{\a_\k}}\nonu
Z^{(u)}&=&\frac{\sqrt{2}}{C_U\a_\k}.\label{allorderZ}
\eea
$W^{(0)}$ was defined in eqs. (\ref{legdef}) and (\ref{renor2}) and the $C$
coefficients were given in eq. (\ref{Cfactors}). These results are fully
consistent with the large $k$ results found in section three.

As discussed, the normalization of the currents, eq. (\ref{Cfactors}), was
computed in the previous section using operator methods.
Choosing for $\a_\k$ the value which is found in the same formalism:
$\a_\k= \k+\frac 1 2 (4-N)$, eq.(\ref{allorderZ}) further simplifies to
\be
Z^{(t)}=Z^{(g)}=Z^{(u)}=\frac{1}{k+N-3}.\label{Zsg}
\ee

Using eqs. (\ref{relsc}) and (\ref{relsd}), we find that the level $\k_c$ of
the $\osp$ affine Lie algebra as a function of the central extension $c$ is
given by:
\bea
12\k_c&=&-c+ 37+ \frac 1 2  N(N-24) \nonu
&& - \left(\left( c - 13 -\frac 1 2 N(N-12)\right)^2  + 6 (N-2)(N-3)(N-4)
\right)^{1/2}.
\label{renormc}
\eea
For $N=0$, one gets real values for $\k_c$ if $0\leq c\leq 1$ or $c\geq 25$;
for $N=1$ the allowed range is $0\leq c\leq 3/2$ or $c\geq 27/2$ and for $N\geq
2$, there is no restriction on the range of $c$.

Furthermore, eq. (\ref{renormc}) implies that while $N=0$ and 1 receive
contributions to the renormalization of $\k_c$ at all loop orders, $N=2$, 3 and
4 receive only one-loop contributions to the renormalization. For $N=4$ an even
stronger statement is possible. The $N=4$ superconformal algebra can be
linearized by adding one $U(1)$ field and four free fermions to the system.
Then
$c$ changes to $c_{\rm lin}=c+3$ and one has $c_{\rm lin}=-6\k$. We conclude
that in this case no renormalization at all occurs and the quantum theory is
equal to the classical one. These results are a reflection of the
non-renormalization theorems for theories with extended supersymmetries.

However, we want to stress here that while the value of the coupling constant
renormalization is unambigously determined, the computation of the value of the
wavefunction renormalization is very delicate. If the gauged WZW model serves
as a guideline \cite{ruud}, we expect that the precise value of the
wavefunction renormalization depends on the chosen regularization scheme.
As mentioned before the computations leading to the quantum effective
action were performed  in the operator formalism using point-splitting
regularization. Within this framework, we believe that eq. (\ref{Zsg}) is fully
consistent. This claim is further supported, as we will show next, by
perturbative computations which also rely on operator methods and which give
results which are fully consistent with both eqs. (\ref{renormc}) and
(\ref{Zsg}).

We want to remark here that computing the $Z$ factors using functional methods
in a certain regularization scheme, probably does not simply amount to eq.
(\ref{allorderZ}) with an appropriate choice for $\a_\k$, as claimed in
\cite{dbg}. This, because the value of $C$, eq. (\ref{Cfactors}), might also
very well depend on the choice of regularization.

\subsection{Semiclassical Evaluation}

In the previous section we computed the renormalisation factors for the
nonlinear $O(N)$ superconformal algebras by realising them as WZW models.
For $N=3$ and $4$, these algebras can be obtained from linear ones
by eliminating the dimension $\half$ fields and for $N=4$ an additional $U(1)$
factor. Also it has been shown
\cite{kris1} that the effective actions $W$ of the linear
theories can be obtained from the linear effective actions  simply
by putting to zero the spin $\half$ currents. In this section we compute
these effective actions for the linear theories in the semiclassical
approximation. A comparison with the results for the nonlinear algebras
obtained so far, eqs.(\ref{Zsg},\ref{renormc}) shows complete agreement
through the linear-nonlinear connection established in \cite{kris1}.

Let us first explain the method\cite{zamo2,stony}.
In the semiclassical approximation, the effective action is computed by a
steepest descent method:
\begin{eqnarray}
e^{-W[u]} &=& \int [dA]e^{-\Gamma [A] - \frac{1}{\pi }\int uA}\nonumber\\
& \simeq & e^{-\Gamma [A_{\rm cl}] - \frac{1}{\pi }uA_{\rm cl}}
\int[d\tilde{A}]
\exp -\frac{1}{2} \tilde A \frac{\delta ^2\Gamma [A_{\rm cl}]}{\delta
A_{\rm cl}\delta A_{\rm cl}}\tilde{A}\,,
\label{effact}
\end{eqnarray}
where $A_{\rm cl}[u]$ is the saddle point value that solves
\begin{equation}
-\frac{\delta \Gamma [A]}{\delta A} = \frac{1}{\pi }u\,, \label{u in A}
\end{equation}
and $\tilde A$ is the fluctuation around this point.  Therefore, all that
has to be done is to compute a determinant:
\begin{equation}
W[u] \simeq W_{\rm cl}[u] + \frac{1}{2} \log \det
  \frac{\delta^2\Gamma [A_{\rm cl}]}{\delta A_{\rm cl}\delta A_{\rm cl}}\,.
  \label{Wsemicl}
\end{equation}
To evaluate this determinant, one may use the Ward identities: schematically,
they have the form
\[\overline{D_1}[A]\frac{\delta \Gamma }{\delta A} \sim \partial _2A\]
where on the l.h.s.\ there is a covariant differential operator, and on the
r.h.s.\ the term resulting from the anomaly, the symbol $\partial _2$
standing for a differential operator of possibly higher order (see for
example eq.~(\ref{wi1}) without the non-linear term).  Taking the
derivative with respect to $A$, and transferring some terms to the r.h.s.\
one obtains
\begin{equation}
\overline{D_1}[A]\frac{\delta ^2\Gamma }{\delta A\delta A} \sim D_2[u]\,,
\label{DG=D}
\end{equation}
where now there appears a covariant operator on the r.h.s. also, with $u$ and
$A$ again related by eq.~(\ref{u in A}).  The sought-after determinant is
then formally the quotient of the determinants of the two covariant
operators in eq.~(\ref{DG=D}).

For the induced and effective actions of fields coupled to affine
currents, the covariant operators are both simply
covariant derivatives, and their determinants are known: both induce a
Wess-Zumino-Witten model action.  For the 2-D
gravity action $A \rightarrow h$ and $u \rightarrow t$, the operator on the
r.h.s.\ is $\partial ^3 + t\partial + \partial t$ and the determinant is in
\cite{zamo2}. The similar computation for the semiclassical
approximation to $W_3$ is in \cite{ssvnc}.  From these cases, one may
infer the general structure of these determinants.
In the gauge where the fields $A$ are fixed, the operator
$\overline{D_1}[A]$ corresponds to the ghost Lagrangian: in BV language,
the relevant piece of the extended action is $A^*\overline{D_1}[A]c +
b^*\lambda $ and as in section four the $A^*$-field is identified
with the Faddeev-Popov antighost. The
determinant is then given by the induced action resulting from the ghost
currents. Since these form the same algebra as the original currents, with
a value of the central extension that can be computed, one has
\be
\log\det\overline{D_1}[A] = k_{ghost}\Gamma ^{(0)}[A]\,.\label{detD1}
\ee
The second determinant can similarly be expressed as a functional integral
over some auxiliary $bc$ and/or $\beta \gamma $ system.  Let us, to be
concrete, take $D_2[h] = \frac{1}{4}(\partial ^3 + t\partial + \partial t)$
as an illustration.  Then we have
\begin{equation}
(\det D_2[t])^{1/2} = \int[d\sigma ] \exp \frac{1}{8\pi }\sigma
(\partial ^3\sigma + t\partial \sigma + \partial (t\sigma ))
\label{hLag}
\end{equation}
where it is sufficient to use a single fermionic integral since $D_2$ is
antisymmetric.  We can rewrite this as
\[(\det D_2[t])^{1/2} = \exp^{-\tilde W[t]} =
\langle \exp \frac{1}{\pi }\int t H \rangle_\sigma \]
where $H = \frac{1}{4}\sigma \partial \sigma $.

The propagator of the $\sigma $-field fluctuations is given by
\be
\sigma (x)\sigma (y) = 2 \frac{(x - y)^2}{\bar x - \bar y}
+ \mbox{regular terms} \label{sigmaprop}
\ee
and the induced action has been called $\tilde W$ instead of $\Gamma $ for
reasons that will be clear soon.\newline
The Lagrangian eq.~(\ref{hLag}) has an invariance:
\begin{eqnarray*}
\delta t &=& \partial ^3 \omega + 2(\partial \omega )t + \omega \partial
t\\
\delta \sigma &=& \omega \partial \sigma - (\partial \omega )\sigma
\end{eqnarray*}
and, correspondingly, $\tilde W[t]$ obeys a Ward identity. This
is anomalous. We find :
\begin{equation}
\partial ^3\frac{\delta \tilde W[t]}{\delta t} + 2t\partial \frac{\delta
\tilde W(t) }{\delta t} + (\partial t)\frac{\partial \tilde W(t)}{\partial
t} = -\frac{1}{\p}\overline{\partial }t\,.
\label{WIB}
\end{equation}
This is nothing but the usual chiral gauge conformal Ward identity `read
backwards', {\it i.e.}\
$t \leftrightarrow \frac{\delta \Gamma [h]}{\delta h} $
and $h \leftrightarrow \frac{\partial \tilde W (t)}{\partial t}$.
We conclude that $\tilde W(t)$ is proportional to the Legendre
transform of $\G^{(0)}$,
\be
\tilde W(t) = -6k' W^{(0)}(t)\label{Wtilde}
\ee
with $k'=2$.
Another way to obtain this Ward identity eq.~(\ref{WIB}) is by evaluating
the operator product of the $H$-operator.  One finds
\begin{equation}
H(z)H(0) = -\frac{k'}{4}\left(\frac{z}{\bz}\right)^2 -
\frac{z}{\bz}H(0) - \frac{z^2}{2\bz}\partial H(0) +
\cdots
\label{Halgebra}
\end{equation}
as in \cite{zamo2}, from which eq.(\ref{Wtilde}) also follows.

 Note that in \cite{zamo2}
different bosonic realisations of the algebra (\ref{Halgebra}) were used.
Starting from the action $\frac{1}{2}\varphi (\partial ^2 +
\frac{t}{2})\varphi $ one finds that $H_\varphi = \frac{1}{4}\varphi ^2$
satisfies (\ref{Halgebra}) with $k' = -1/2$. This realisation will appear
naturally when we discuss $N=1$. Another one starting from
$\varphi _1[\partial ^3 + t\partial + \partial t]\varphi _2$ has $k' = -4$
and is a bosonic twin of the one we used.
The same algebra also realises a connection with $sl_2$,
through (\cite{zamo2}) $H(z) =
-\frac{z^2}{2} j^+(\bz) + zj^0(\bz) +
\frac{1}{2}j^-(\bz)$: The antiholomorphic components $j^a$ of
$H(z,\bz)$ generate an affine $sl_2$ algebra.

The upshot is that whereas the first determinant is proportional to the
classical induced action $\Gamma$, the second one is proportional to
the classical effective action $W$.  The proportionality constants
are pure numbers independent of the central extension of the original
action.  From these numbers the renormalisation factors for the quantum
effective action in the semiclassical approximation follow:
\begin{eqnarray*}
W[u] &\simeq& kW_d^{(0)}\left[\frac{u}{k}\right]
-6k' W^{(0)}\left[\frac{u}{k}\right] - \frac{k_{ghost}}{2}\Gamma ^{(0)}[A] \\
&\simeq& \left(k -6k' -
\frac{k_{ghost}}{2}\right)W_d^{(0)}\left[\frac{u}{k}\right] +
\frac{k_{ghost}}{2}u\frac{\partial W_d^{(0)}}{\partial
u}\left[\frac{u}{k}\right]\\
&\simeq& \left(k -6k' -
\frac{k_{ghost}}{2}\right)W_d^{(0)}\left[\frac{u}{k}(1 +
\frac{k_{ghost}}{2k})\right]\,. \label{Wusemicl}
\end{eqnarray*}
In the example of affine KM-currents, the relevant numbers are
$6k' =- \tilde h$ and $k_{ghost} = -2\tilde h$, with the familiar result
\footnote{Different calculations give different answers for the field
renormalisation factor. See the discussion at the end of the previous
subsection, appendix B, and \cite{ruud}}.
For $W_2$-gravity, the results of \cite{zamo2} follow.  We now turn to $N
= 1 \cdots 4$ linear supergravities.\newline
\fbox{$N = 1$}\newline
This case has been treated also in \cite{zamo2,PawMeiss}.
The induced action is
\[e^{-\Gamma [h,\psi ]} = \langle e^{-\frac{1}{\pi }(hT + \psi G)}\rangle\]
where $T$ and $G$ generate the $N = 1$ superconformal algebra with central
charge $c$.  The Ward identities read\footnote{All derivatives are left
derivatives.}, with $\Gamma = c\Gamma ^{(0)}$,
\begin{eqnarray}
[\overline{\partial } - h\partial - 2(\partial h)]\frac{\delta \Gamma
^{(0)}}{\delta h} - \frac{1}{2}[\psi \partial - 3(\partial \psi
)]\frac{\delta \Gamma ^{(0)} }{\delta \psi } &=& \frac{1}{12\pi }\partial
^3 h \nonumber \\
{[}\overline{\partial } - h\partial - \frac{3}{2}(\partial h)]\frac{\delta
\Gamma ^{(0)} }{\delta \psi } - \frac{1}{2}\psi \frac{\delta \Gamma
^{(0)}}{\delta h} &=& \frac{1}{3\pi } \partial ^2\psi\,.
\label{WI1}
\end{eqnarray}
{}From this we read off $\bar{D_1}$ and $D_2$ of eq.~(\ref{DG=D}):
\begin{eqnarray*}
\overline{D_1} &=& \left(\begin{array}{cc}
\overline{\partial} - h\partial - 2(\partial h) & -\frac{1}{2}\psi \partial
+ \frac{3}{2}(\partial \psi )\\
-\frac{1}{2}\psi  & \overline{\partial } - h\partial - \frac{3}{2}(\partial
h)\end{array}\right)\\
D_2 &=& \frac{1}{3\pi }\left(\begin{array}{cc}
\frac{1}{4}(\partial ^3 + (\partial \hat t) + 2\hat t\partial ) &
-\frac{1}{2} (\partial \hat y) - \frac{3}{2}\hat g \partial \\
(\partial \hat g) + \frac{3}{2}\hat g\partial  & \partial ^2 + \hat
t/2\end{array}\right)
\end{eqnarray*}

 We abbreviated $\hat t = -12\pi \frac{\partial \Gamma ^{(0)}[4]}{\partial h} =
t/c$ and $\hat g = g/c$.  For ease of notation, we will drop the hats in
the computation of $\det D_2$.

  $\overline{D_1}$ gives rise to the
ghost-realisation for $N = 1$, so we have
\[ \mbox{s}\det \overline{D_1} = 15\Gamma ^{(0)}(h,\psi ).\]
$D_2$ is a (super)antisymmetric operator, as can be seen by rewriting
\begin{eqnarray*}
\partial ^3 + (\partial t) + 2 t \partial &=& \partial ^3 +
\partial t + t\partial \,,\\
\frac{1}{2}(\partial g) + \frac{3}{2} g \partial  &=&
\frac{1}{2}\partial g + g \partial \,,\\
(\partial g) + \frac{3}{2} g \partial  &=& \frac{1}{2} g
\partial  + \partial g\,.
\end{eqnarray*}
The relevant action is then
\[\frac{1}{\pi } \int\left(\frac{1}{8}\sigma (\partial + \partial t +
t\partial )\sigma  - \frac{1}{2}\sigma (\partial g)\varphi - \frac{3}{2}
\sigma g\partial \varphi + \frac{1}{2}\varphi \left(\partial ^2 +
\frac{t}{2}\right)\varphi \right)\,.\]
The determinant is
\begin{equation}
(s \det D_2)^{1/2} = \langle e^{-\frac{1}{\pi }(tH + g\Psi )}\rangle =
e^{-\tilde W(t,g)}
\label{det2}
\end{equation}
where $H = \frac{1}{4}\sigma \partial \sigma + \frac{1}{4}\varphi ^2$ and
$\Psi = -\frac{1}{2}(\partial \sigma )\varphi + \sigma \partial \varphi $
and the average is taken in a free field sense with propagators
eq.(\ref{sigmaprop}) and
\begin{eqnarray}
\langle \varphi (x)\varphi (y)\rangle &=& \frac{ x - y}{\bar x - \bar y}\,.
\label{phiprop}
\end{eqnarray}
This leads to the operator product expansions
\begin{eqnarray}
H(z)H(0) &=& -\frac{k'}{4}\frac{z^2}{\bz^2} +
\frac{z}{\bz}H(0) + \frac{1}{2}\frac{z^2}{\bz}\partial
H(0) + \cdots\nonumber\\
H(z)\Psi (0) &=& \frac{1}{2}\frac{z}{\bz}\Psi (0) +
\frac{1}{2}\frac{z^2}{\bz}\partial \psi (0) + \cdots\nonumber\\
\Psi (z)\Psi (0) &=& 2k' \frac{z}{\bz^2} -
\frac{4}{\bz}H(0) - \frac{2z}{\bz}\partial H(0) + \ldots
\label{HPSIalgebra}
\end{eqnarray}
with a value for the central extension $k' = 2 - \frac{1}{2}$.\newline
The resulting Ward identities for $\tilde W$, eq.~(\ref{det2}), are:
\begin{eqnarray*}
(\partial ^3 + (\partial t) + 2t\partial )\frac{\partial \tilde
W}{\partial t} -(2(\del g)+6g\del) \frac{\partial \tilde W}{\partial g} &=&
-\frac{k}{2\pi }\overline{\partial }t\\
(\partial ^2+\frac{t}{2}) \frac{\partial \tilde W}{\partial g}
 +(\,\,(\del g)+\frac{3}{2}g\del) \frac{\partial \tilde W
}{\partial t}&=& -\frac{2k}{\pi }\overline{\partial }g\,.
\end{eqnarray*}
Comparing with eq.~(\ref{WI1}), and reverting to the proper normalisation
of $t$ and $g$, we have
\[\tilde W(\hat t,\hat g) =
   6k'\ W^{(0)}[\hat t,\hat g] = 6k'\ W^{(0)}[t/c,g/c]\]
where we used
\[W^{(0)}(\hat t,\hat g) = \min_{\{h,\psi \}}\left(\Gamma ^{(0)}(h,\psi ) -
\frac{1}{12\pi }\int h\hat t - \frac{1}{3\pi }\int \hat g \psi \right)\,.\]
Putting everything together in eq.~(\ref{Wsemicl}) we find, for $N = 1$,
$k' = 3/2$,
\[W[t,g] \simeq c W_d^{(0)}\left[\frac{t}{c},\frac{g}{c}\right] -
\frac{15}{2}\Gamma ^{(0)}[h,\psi ] -
9 W^{(0)}\left[\frac{t}{c},\frac{g}{c}\right]\,.\]
Writing these results as
\[W^{(N)}[\Phi ] \simeq Z_W^{(N)}  W^{(0)}[Z_\Phi ^{(N)}\Phi ]\]
we have
\begin{eqnarray*}
&&Z_W^{(1)} \simeq c - \frac{33}{2}\\
&&Z_t^{(1)} = Z_g^{(1)} \simeq \frac{1}{c}(1 + \frac{15}{2c})\,.
\end{eqnarray*}
For reference, the corresponding equations for $N = 0$ are $k' = 2$,
\begin{eqnarray*}
W[t] &=& c W_{cl}^{(0)} [t/c] - \frac{26}{2} \Gamma ^{(0)}[h] - 12 W^{(0)}
[t/c]\\
Z^{(0)}_W &=& c - 25\\
Z^{(0)}_t &=& \frac{1}{c}(1 + \frac{13}{c}).
\end{eqnarray*}
These values are in complete agreement with
\cite{zamo2,kpz,sufrac,PawMeiss} and with our section 5.

Before going to $N=2$, we comment on the technique we used to
obtain eq. (\ref{Halgebra}) and (\ref {HPSIalgebra}). The easiest way is
expand the fields $\sigma ,\varphi $ in solutions of the free field
equations:
\begin{eqnarray*}
\sigma (z,\bz) &=& \sigma ^{(0)} (\bz) + z \sigma ^{(1)} (\bz)
+ \frac{z^2}{2} \sigma ^{(2)} (\bz)  \\
\J(z,\bz) &=& \J^{(0)} (\bz) + z \J^{(1)} (\bz)
\end{eqnarray*}
and read off the OPE's for the antiholomorphic coefficients from eqs.
(\ref{sigmaprop},\ref{phiprop}). Then all singular terms are given in eq.
(\ref{HPSIalgebra}).
An alternative would be, to use Wick's method, with the contractions given
by the propagators. The resulting bilocals then give, upon Taylor-expanding
the same algebra as in eq.(\ref{HPSIalgebra}), up to terms proportional
to equations of motion. This ambiguity was already present in
\cite{zamo2}, see also \cite{ssvnc}.
We have simply used the antiholomorphic mode expansion in
the following calculation. A disadvantage is, that in this way one loses
control over equation of motion terms.

Let us close the $N=1$ case by noting that, just as the antiholomorphic
modes corresponding to eq. (\ref{Halgebra}) generate on $sl(2)$ affine
algebra, we get an affine $osp (1|2)$ from the modes of $H$ and $\J$ of
eq. (\ref{HPSIalgebra}).\newline
\fbox{$N = 2$}\newline
For $N=2$ the extension of the scheme above has two $\J$-fields and a
free fermion $\tau$, with $<\tau(x)\tau(0)> = - \frac{1}{\bar{x}}$. This
last field does not contribute to $H$:
\begin{eqnarray*}
H    &=& \frac{1}{4} \sigma \partial \sigma
                  + \frac{1}{4} \sum^2_1 \phi^2_a,                    \\
\J_a &=& - \frac{1}{2} (\partial \sigma ) \phi_a
                + \sigma  \partial \phi_a - \varepsilon _{ab}\phi_b \tau, \\
A    &=& \varepsilon _{ab} \partial \phi_a \phi_b+\sigma \del \tau.
\end{eqnarray*}
Note that $\del A$ is proportional to the equations of motion for $\phi_a$
and $\tau$.
Neglecting terms proportional to equations of motion, we find that
in the algebra of eq. (\ref{HPSIalgebra}) the first two equations are
supplemented with
\begin{eqnarray}
H(z)A(0) &=& 0 + \cdots\nonumber\\
\Psi _a(z)\Psi _b(0) &=& \delta _{ab}\left(\frac{2k'z}{\bz^2} -
\frac{4H(0)}{\bz} - \frac{2z}{\bz}\partial H(0)\right) +
\varepsilon _{ab} \frac{z}{\bz}A(0) + \cdots\nonumber\\
A(z)\Psi _a(0) &=& \frac{\varepsilon _{ab}}{\bz}\Psi _b(0) +
\cdots\nonumber\\
A(z)A(0) &=& \frac{k'}{\bz^2} + \cdots\,,
\label{HPSI2 algebra}
\end{eqnarray}
and $k' = 2 - 2\cdot \frac{1}{2} = 1$.  With the central charge of the
ghosts being $k_{ghost} = +6$, we can immediately write down the
$Z$-factors for $N=2$:
\begin{eqnarray*}
Z_W^{(2)} &=& c - 9\\
Z_t^{(2)} &=& Z_g^{(2)} = Z_a \simeq \frac{1}{c}(1 + 3/c)\,.
\end{eqnarray*}
The algebra of antiholomorphic coefficients of $H,\J_a$ and $U$ is now
$osp(2|2)$.

It should be remarked that for $N = 2$ (and higher) the algebra
eq.~(\ref{HPSI2 algebra}) does not quite reproduce the Ward identities for
the induced action.  Here also, the equations of motion are involved.  The
difference is in the identity.
\[\frac{1}{4}(\partial ^3 + 2t\partial + (\partial t))h -
\frac{1}{2}((\partial g_a) + 3g_a\partial )\psi ^a - (\partial A)\cdot u =
\overline{\partial }t\,.\]
The last term, as noted, is proportional to equations of motion of the
free part of the action of the auxiliary system, and is not be recovered
from the procedure outlined above.
 We surmise that, as for $N = 0$ and $1$,
 these terms do not change the result.
\newline
\fbox{$N = 3,4$}\newline
The determinant leads to consider the Lagrangian
\begin{eqnarray*}
L &=& \frac{1}{8}\sigma \partial ^3\sigma + \frac{t}{4}\sigma \partial
\sigma - \frac{1}{2}\sigma \partial g\cdot \varphi - \frac{3}{2}\sigma
g\cdot \partial \varphi + \sigma u \cdot \partial \tau - \frac{\sigma
}{2}(q\partial \overline{q} - \partial q\overline{q})\\
&& \frac{1}{2}\varphi \cdot \partial ^2\varphi  + \frac{t}{4}\varphi \cdot
\varphi - u \cdot \varphi \wedge \partial \varphi + \varphi
\cdot g\wedge \tau  - q\varphi \cdot \partial \tau + u \cdot \varphi
\overline{q}\\
&&- \frac{1}{2}\tau \cdot \partial \tau + \frac{1}{2}u \cdot \tau \wedge
\tau  - \frac{1}{2}\overline{q}^2\\
&=& \frac{1}{8}\sigma \partial ^3\sigma  + \frac{1}{2}\varphi  \cdot
\partial ^2\varphi - \frac{1}{2}\tau \partial \tau  -
\frac{1}{2}\overline{q}^2+ tH + g \cdot \Psi + u \cdot A + q \cdot \Theta
\end{eqnarray*}
where $\sigma ,t,q$ and $\overline{q}$ are $O(3)$ scalars and $g,\varphi
,u,\tau $ are $O(3)$ vectors.  It has the following invariances, with
scalar parameters $\omega $ and $\beta$ and vectors $\theta $ and $\alpha
$: \begin{itemize}
\item $\delta t = \partial ^3\omega + \omega \partial t + 2(\partial \omega
)t$\\*[3mm]
$\delta \chi = \omega \partial \chi + j(\partial \omega )\chi$ \\*[3mm]
with $j = {\displaystyle
\frac{3}{2},1,\frac{1}{2},-1,-\frac{1}{2},0,\frac{1}{2}}$ for $\chi =
g^a,u^a,q,\sigma ,\varphi ^a,\tau ^a,\overline{q}$
\item \parbox[t]{7cm}{$\delta t = -2(3\partial \theta \cdot g + \theta
\cdot \partial g)$\\*[3mm]
$\delta g = \partial ^2\theta - \theta \wedge \partial u - 2\partial \theta
\wedge u + {\displaystyle \frac{t}{2}}\theta $\\*[3mm]
$\delta u = -\theta \wedge g + \partial q \theta + q \partial \theta
$\\*[3mm]
$\delta q = \theta u$}\ \ \parbox[t]{7cm}{$\delta \sigma  = 2\theta \cdot
\varphi $\\*[3mm]
$\delta \varphi = -\theta \wedge \tau  - {\displaystyle \frac{1}{2}}\theta
\partial \sigma + \partial \theta \sigma $\\*[3mm]
$\delta \tau = -\varphi \wedge \partial \theta + \partial \varphi \wedge
\theta + \theta \overline q$\\*[3mm]
$\delta \overline{q} = -\theta \partial \tau $}
\item \parbox[t]{7cm}{$\delta t = -4u\cdot \partial \alpha $\\*[3mm]
$\delta g = g \wedge \alpha - q \partial \alpha $\\*[3mm]
$\delta u = \partial \alpha + u \wedge \alpha $\\*[3mm]
$\delta q = 0$}\ \ \parbox[t]{7cm}{$\delta \sigma  = 0$\\*[3mm]
$\delta \varphi = \varphi \wedge \alpha $\\*[3mm]
$\delta \tau = \sigma \partial \alpha + \tau \wedge \alpha $\\*[3mm]
$\delta \overline{q} = \varphi \partial \alpha $}
\item \parbox[t]{7cm}{$\delta t = -2q\partial \beta + 2(\partial q)\beta
$\\*[3mm]
$\delta g = -u\beta $\\*[3mm]
$\delta u = 0$\\*[3mm]
$\delta q = \beta $}\ \ \parbox[t]{7cm}{$\delta \sigma  =0$\\*[3mm]
$\delta \varphi =0$\\*[3mm]
$\delta \tau =-\varphi \beta$\\*[3mm]
$\delta \overline{q} =+\frac{1}{2}\partial \sigma \beta+\sigma
\partial\beta$} \end{itemize}
Note that $\overline{q}$ is an auxiliary field. Again working up
to equations of motion as before, one finds that $H,\Psi ,A$ and $\Theta $
form a closed operator algebra that extends eq.(\ref{HPSI2 algebra})
straightforwardly.  The resulting anomalous Ward identities again afford
the conclusion that the functional determinant is proportional to the
appropriate effective action.  For $N = 4$, we refrain from writing out the
action and transformation laws, but the same results are valid.
The algebra
of eq.~(\ref{HPSI2 algebra}) only changes in that more $\Psi _a$ and
$A$ fields are present.  The value of the central charge $k'^{(N)}$ in that
algebra can most simply be obtained from $H(z)H(0)$, since only $\sigma $
and $\varphi _a$ fields contribute to it:
\[k'^{(N)} = 2 - \frac{N}{2}\,.\]
The resulting antiholomorphic coefficients constitute an affinisation of
the $osp(N|2)$ superalgebra: the spin $\half$ field $\Theta$
contributes no antiholomorphic modes.

The ghost system central charges vanish for $N = 3,4$.  As a result, for $N
= 3$,
\begin{eqnarray*}
Z_W^{(3)} &=& c - 3,\\
Z_t^{(3)} &\simeq& \frac{1}{c},
\end{eqnarray*}
and for $N = 4$, all $Z$-factors are equal to their classical
values.

Now we compare these results for $Z_W$ with the results
of eqs.~(\ref{allorderZ},\ref{renormc}) for the {\it nonlinear} algebras,
using the result of \cite{kris1}.
According to \cite{kris1}, the respective effective
actions are equal upon putting the appropriate currents to zero.
Recall that the
linear algebras reduce to the nonlinear ones when eliminating
\cite{goddard} one spin $1/2$ field for $N=3$, and four spin $1/2$
fields and one spin $1$ field for $N=4$. In the process, the central charge
is modified: \begin{eqnarray*}
c_{nonlinear}^{(3)} &=& c_{linear}^{(3)} - 1/2\\
c_{nonlinear}^{(4)} &=& c_{linear}^{(4)} - 3\,.
\end{eqnarray*}
With these substitutions, the agreement with eq.~({\ref{renormc}) is
complete, both for the overall renormalisation factor and for the field
renormalisations.

For $N=3$ a similar computation was made \cite{gustav} directly on the
 theory based on the nonlinear algebra, using Feynman diagrams to compute
the determinants. In that case the classical approximation is not linear in
$c$, but can be written as a power series. The determinant replacing our
$\mbox{s}\det \overline{D_1}$ is not directly proportional to the induced
action $\G^{(0)}$ as in eq.(\ref{detD1}).
In fact, this part vanishes since $k_{ghost}=0$ for $N=3$.
Instead, it contains extra terms. These terms are computed in
\cite{gustav}. They cancel the non-leading terms of the classical
induced action, at least to the extent they are relevant here
(next-to-leading order).  A similar cancellation was also observed in the
computation of the $W_3$ effective action \cite{ssvnb}.
The non-leading contribution and its cancellation with some of the loop
contributions seems to have been overlooked in \cite{gustav}. We have
recomputed the renormalisation factors for the nonlinear algebra
with the method of \cite{gustav}, taking into account the non-leading terms
also. We again find agreement with the results obtained above. Note in
particular that all field renormalisation factors are equal. This
alternative computation of the determinants, using Feynmann diagrams,
implicitely confirms our treatment of equation of motion terms in the
Ward identities.

\section{The General Case}
\setcounter{equation}{0}
\setcounter{footnote}{0}

Given a (super) affine Lie algebra $\hat{g}$ of level $\k$, we call the finite
dimensional  subalgebra $\bar{g}$. To every inequivalent, nontrivial embedding
of $\slt$ in $\bar{g}$ there corresponds a certain extension of $d=2$
gravity\footnote{This is merely rephrasing the fact that to every inequivalent,
nontrivial embedding of $\slt$ in $\bar{g}$, one can associate an extension of
the Virasoro algebra.}. In this section we will develop the general procedure
to construct the effective theory. Details about $\slt$ embeddings can be found
in appendix C.

A nontrivial embedding of $\slt$ in $\bar{g}$ is given. In \cite{bais}, it was
shown that by constraining $\del g g ^{-1}$ as
\be
\del g g ^{-1}=e_-+u_z,\label{aastd}
\ee
where
\be
u_z\in \mbox{Ker ad}\,e_+,
\ee
the flat connection condition reduces to the large $\k$ limit of the Ward
identities of an extension of the Virasoro algebra. For every spin
$j$ $\slt$
multiplet one has a current with conformal dimension given by $j+1$.
We will now implement these constraints through a gauged WZW model.

One easily checks that
\be
\dim\, \mbox{Ker ad}\,e_+=\left( \dim\,\bar{g}+\dim\P_{+1/2 }\bar{g}\right)-2
\left( \dim\P_{\geq +1 }\bar{g}+\dim\P_{+1/2 }\bar{g}
\right),
\ee
which,
in the case that $\P_{+1/2}\bar{g}\neq\emptyset$, strongly suggests the need to
introduce $\dim\P_{+1/2 }\bar{g}$ auxiliary fields $\t\in\P_{-1/2 }\bar{g}$.
When $\bar{g}$ is a purely bosonic algebra, it was shown in {\it e.g.}
\cite{dublin,alex} that there is no need for the auxiliary fields. This is
due to the fact that in that case  $\dim\P_{+1/2 }\bar{g}$ is always even. The
branching always contains a $U(1)$ generator under which the multiplets with
$j$ halfinteger can be split into two subspaces having opposite eigenvalues
under the action of the $U(1)$ symmetry. Instead of taking $\P_+\bar{g}$ as the
gauge group, one splits $\P_{+1/2}\bar{g}$ according to the $U(1)$ chirality as
$\P^{B}_{+1/2}\bar{g}=\P^{+B}_{+1/2}\bar{g}+\P^{-B}_{+1/2}\bar{g}$ and one
takes $\P^{B}_{\geq +1}\bar{g}+\P^{+B}_{+1/2}\bar{g}$ as gauge group. The
other constraints are then regained as gauge fixing conditions. As the
example of $N=1$ supergravity already shows, the introduction of extra fields
is in certain cases unavoidable\footnote{Note that even in the supersymmetric
case, auxiliary fields can sometimes be avoided. One example is $N=2$
supergravity where the $U(1)$ symmetry can be used to restrict the gauge
group.}. In order to give a unified description, we always introduce extra
fields whenever representations of half-integral dimension occur.

Introduce the gauge fields $A_{\bz}\in\P_{+}\bar{g}$ and the ``auxiliary''
fields $\t\in\P^F_{-1/2}\bar{g}$, $r\in\P^{+B}_{-1/2}\bar{g}$ and
$\bar{r}\in\P^{-B}_{-1/2}\bar{g}$ where we denote
\bea
r&\equiv&\sum_{j,\a}r^{(j,\a,\b=1)}t_{(j\,-1/2,\a)}\nonu
\bar{r}&\equiv&\sum_{j,\a,\b=-1}\bar{r}^{(j,\a)}\bar{t}_{(j\,-1/2,\a)}\nonu
\t&\equiv&\sum_{j',\a'}\t^{(j'\a')}\hat{t}_{(j'\,-1/2,\a')},
\eea
and $t_{(j\,-1/2,\a)}\in\P^{+B}_{-1/2}\bar{g}$,
$\bar{t}_{(j\,-1/2,\a)}\in\P^{-B}_{-1/2}\bar{g}$ and
$\hat{t}_{(j'\,-1/2,\a')}\in\P^{F}_{-1/2}\bar{g}$.

Using elementary properties of $\slt$ representations and some of the results
in appendix C, one shows that the action ${\cal S}_1$:
\bea
{\cal S}_1&=&\k S^-[g]+ \frac{\k}{2\p x} \int str\, A_{\bz}\left( \del g g^{-1}
-e_- -r-\bar{r}-\t\right)\nonu
&&- \sum_{j'\a'}\frac{\k y}{(j'+\frac 1 2)\p}\int \t^{(j',\a')}\bdel
\t^{(j',\a')}
- \sum_{j\a}\frac{2\k y}{(j+\frac 1 2 )\p}\int \bar{r}^{(j,\a)}\bdel
r^{(j,\a)},
\label{actiongen}
\eea
is indeed invariant under gauge transformations with
parameters $\P_+\bar{g}$, provided $A_{\bz}$, $r$, $\bar{r}$ and $\t$
transform as \bea
\d A_{\bz}&=&\bdel\h +[\h,A_{\bz}]\nonu
\d r&=&\left[  \P^{+B}_{+1/2}\h,e_- \right]\nonu
\d \bar{r}&=&\left[  \P^{-B}_{+1/2}\h,e_- \right]\nonu
\d \t&=&\left[  \P^{F}_{+1/2}\h,e_- \right]
\eea
and $\h\in\P_+\bar{g}$.

Using exactly the same methods as in section 5, we can construct the
polynomials $T\equiv\sum_{j,\a_j}T^{(j,\a_j)}t_{(jj,\a_j)}$ which are gauge
invariant modulo the field equations of the gauge fields:
\be
T\equiv  \mbox{e}^{\dis\ln\sqrt{C} e_0}\P_{\ker{\rm ad}e_+}J_z \mbox{e}^{\dis -
\ln\sqrt{C} e_0}+\cdots\label{Tgen}
\ee
and the normalization constant $C$ is determined by the {\it requirement} that
$T^{(1,0)}$ generates the Virasoro algebra\footnote{Note that this
normalization differs by constant factors from the one used in the study of
$SO(N)$ supergravity.}. We couple these currents to sources and modify the
action to
\be
{\cal S}_2={\cal S}_1+\frac{1}{4\p x y} \int str h T,
\ee
where the sources $h$ are given by
\be
h\equiv \sum_{j,\a_j}h_{(j,\a_j)}t_{(j\,-j,\a_j)}.
\ee
The action ${\cal S}_2$ is gauge invariant provided we modify the
transformation rules for the gauge fields suitably.
These modifications are proportional to the $h$-fields. One then proceeds by
solving the BV master equation. This will introduce ghost fields
$c\in\P_+\bar{g}$. Though we cannot give the solution of the BV master equation
 in its full generality, we have enough information to proceed and choose a
gauge. The chosen gauge corresponds to putting $A_{\bz}=0$ and will allow us to
determine the normalization of the currents and the value of the central
extension $c$. As in section 4, this gauge choice is accomplished by changing
$A^*_{\bz}$ into a field,
the antighost $b\in\P_-\bar{g}$, and $A_{\bz}$ into an antifield $b^*$.
The action ${\cal S}_1$, eq. (\ref{actiongen}), together with its
gauge transformation rules, is just sufficient to determine the gauge fixed
action to be
\bea
{\cal S}_{\rm gf}&=&\k S^-[g]
- \sum_{j'\a'}\frac{\k y}{(j'+\frac 1 2 )\p}\int \t^{(j',\a')}\bdel
\t^{(j',\a')}
- \sum_{j\a}\frac{2\k y}{(j+ \frac 1 2 )\p}\int \bar{r}^{(j,\a)}\bdel
r^{(j,\a)}\nonu
&&+\frac{1}{2\p x}\int str\, b\bdel c+\frac{1}{4\p x y} \int str h \hat{T},
\label{actiongen2}
\eea
where $\hat{T}$ will be discussed shortly.
The BRST charge also follows, since the transformation laws of all relevant
fields are known, including the $b$-ghosts which are known explicitly from
the term proportional to $A_{\bz}$ in eq. (\ref{actiongen}):
\be
Q=\frac{1}{4\p i x}\oint str\left\{ c \left( J_z -\frac \k 2
\left(e_-+\t+r+\bar{r}\right)\right)+ \frac 1 2 b c c \right\}.
\ee
It is nilpotent.

For $h=0$ the action is BRST invariant. In order to guarantee BRST invariance
for $h\neq 0$, the currents $\hat{T}$ themselves have to be BRST invariant.
For these currents we have not given an explicit form. In
fact they absorb in the present gauge all complications arising from the
non-closure terms in the extended action. However, the requirement of BRST
invariance determines them up to BRST exact pieces. As we will see next, even
this ambiguity  can be eliminated by considering a reduced BRST complex.

We now study the BRST cohomology in some detail. We will use
methods inspired by \cite{tjin1,tjin2} and without explicitly  mentioning,
several results from \cite{bott}. However the presence of the auxiliary fields
$\{ \t,r,\bar{r}\}$ considerably complicates the analysis. Our arguments will
be somewhat heuristic and we postpone a rigorous derivation to a future
publication.

We split the BRST charge into three parts $Q=Q_0+Q_1+Q_2$, where
\bea
Q_0&=&-\frac{\k}{8\p i x}\oint str c e_-\nonu
Q_1&=&-\frac{\k}{8\p i x}\oint str c\left(\t+r+\bar{r}\right).\label{qsqr}
\eea
One has $Q_0^2=Q_2^2=\{Q_0,Q_1\}=\{Q_1,Q_2\}=0$ and
\be
Q_1^2=-\{Q_0,Q_2\}=\frac{\k}{32 \p i x}\oint str\left\{ c \left[
\P_{1/2}c,e_-\right] \right\}.\label{Q12}
\ee
The action of $Q=Q_0$, $Q_1$ and $Q_2$ on the basic fields is given by
{\small
\be
\begin{array}{rcclrcclrccl}
Q_0\,: &  b & \rightarrow & -\frac \k 2 e_- &
Q_1\,: &  b & \rightarrow & -\frac \k 2 (\t+r+\bar{r}) &
Q_2\,: &  b & \rightarrow & \P_-\hat{J}_z\\
 &c &\rightarrow&  0&
 &c &\rightarrow & 0&
 &c &\rightarrow & \frac 1 2 cc  \\
 &\hat{J}_z& \rightarrow&-\frac \k 4 [e_-,c]&
 &\hat{J}_z& \rightarrow&-\frac \k 4 [\t+r+\bar{r},c]&
 &\hat{J}_z& \rightarrow&\frac 1 2 \vec{[}c,\overline{\P}_-\hat{J}_z]+\frac \k
4 \del c \\
 & & & & & & & & & & &+\frac 1 2 [\P_-(t^a),[\P_+(t_a),\del c ]] \\
 &r &\rightarrow&  0&
 &r &\rightarrow & \frac 1 2 [\P_{+1/2}^{+B}c,e_-]&
 &r &\rightarrow & 0  \\
 &\bar{r} &\rightarrow&  0&
 &\bar{r} &\rightarrow & \frac 1 2 [\P_{+1/2}^{-B}c,e_-]&
 &\bar{r} &\rightarrow & 0  \\
 &\t &\rightarrow&  0&
 &\t &\rightarrow & \frac 1 2 [\P_{+1/2}^{F}c,e_-]&
 &\t &\rightarrow & 0,
\end{array}\label{qonf}
\ee
}
where
\be
\hat{J}_z\equiv J_z + \frac 1 2 \{b,c\}
\ee
and $\overline{\P}_-\equiv {\bf 1}-\P_-$ and $\vec{[}A,B]$ stands for
\be
\vec{[}A,B]=(-)^{(ab)}\left(A^aB^b \right)f_{ab}{}^ct_c,
\ee
where $\left(A^aB^b \right)$ is a regularized product.

Consider the algebra $\ca$ generated by the basic fields
$\{b,\hat{J}_z,\t,r,\bar{r},c\}$, which consists of all regularized
products\footnote{We use the standard point-splitting regularization: $(AB)(z)=
\frac{1}{2\p i}\oint dz'(z'-z)^{-1}A(z')B(z)$.} of the basic fields and their
derivatives modulo the usual relations
\cite{bbss,spindel}
between different
orderings, derivatives, etc. We assign ghostnumber $-1$ to $b$, $1$ to $c$ and
$0$ to all other fields. Through this $\ca$ acquires a grading
$\ca=\bigoplus_{n\in{\bf Z}}\ca_n$ where the elements of $\ca_n$ have
ghostnumber $n$. The BRST charge $Q$ is a map from $\ca_n$ to $\ca_{n+1}$ and
it acts as a derivation on a regularized product of fields. In view of the
application at hand, we want to study the cohomology of this complex at
ghostnumber zero, $H^0(\ca )$.

The cohomology of the subcomplex $\ca^{(1)}$, generated by $\{ b,
\P_-\hat{J}_z-\frac \k 2 ( \t+ r+\bar{r})\}$ is $H^*(\ca^{(1)})={\bf C}$. As
such, we have that $H^*(\ca)=H^*(\wha)$ where $\wha=\ca / \ca^{(1)}$ which we
choose to be generated by $\{\overline{\P}_-\hat{J}_z,\t,r,\bar{r},c\}$.

Next we observe that $\wha$ has yet another subcomplex $\ca^{(2)}$ generated by
 $\{\t,r,\bar{r},\P_{1/2}c\}$. Its cohomology is $H^*(\ca^{(2)})={\bf C}$.
Though we have that $H^*(\wha)=H^*(\wta)$ where $\wta=\wha/\ca^{(2)}$, it is
not straightforward to find a set of generators for $\wta$ on which the action
of $Q$ closes in the strong sense. In other words we need to find fields
$\overline{\P}_-\tilde{J}_z$ and $\P_{>1/2}\tilde{c}$ which generate the
algebra $\wta$ and on which the action of $Q$ closes. In order to do this we
introduce $\t'$, $r'$ and $\bar{r}'$:
\bea
\t'&=&\sum_{j,\a}(-)^{j+\frac 1 2}\frac{2}{j+\frac 1
2}\t^{(j,\a)}t_{(j\frac 1 2 , \a)}\nonu
r'&=&\sum_{j,\a,\b=1}(-)^{j+\frac 1 2}\frac{2}{j+\frac 1 2}r^{(j,\a)}
t_{(j\frac 1 2 , \a)}\nonu
\bar{r}'&=&\sum_{j,\a,\b=-1}(-)^{j+\frac 1 2}\frac{2}{j+\frac 1 2}
\bar{r}^{(j,\a)}t_{(j\frac 1 2 , \a)}.
\eea
These fields transform in a very simple way under $Q$:
$\t'\rightarrow\P_{1/2}^{F}c$, $r'\rightarrow\P_{1/2}^{+B}c$ and
$\bar{r}'\rightarrow\P_{1/2}^{-B}c$. Using this, one can now recursively
construct $\P_{>1/2}\tilde{c}$. One finds e.g. $\P_{1}\tilde{c}=\P_1c -\frac 1
4 \mbox{ad}(\t'+r'+\bar{r}')\P_{1/2}c$, $\P_{3/2} \tilde{c}=\P_{3/2} c-\frac 1
2 \mbox{ad} (\t'+r'+\bar{r}')\P_{1}\tilde{c}-\frac{1}{24}
(\mbox{ad}(\t'+r'+\bar{r}'))^2\P_{1/2}c$, etc... So in general, we will have
\be
\P_m\tilde{c}=\sum_{n=0}^{2m-1}(-)^na^{(m)}_n\left(\mbox{ad}(\t'+r'+\bar{r}
' ) \right)^n\P_{m-n/2}c,
\ee
where $a^{(m)}_0=1$, and one has {\it e.g.} $a^{(2)}_1=1/2$,
$a^{(2)}_2=1/8$ and $a^{(2)}_3=1/64$. One proceeds in a similar manner
for the construction of $\overline{\P}_-\tilde{J}_z$. One finds
{\it e.g.} $\P_0\tilde{J}_z=\P_0\hat{J}_z+\frac \k 8[\t,\t']$.

On the reduced complex $\wta$ we get from eq. (\ref{Q12}) that $Q_0$ and
$Q_1+Q_2$ yield a double complex in a weak sense.
We conjecture that using these observations one can show that the
cohomology of $Q$ is isomorphic to that of $Q_0$ on $\wta$. The cohomology
of $Q_0$ is given by $\{\hat{J}^{\rm HW}_z\}=\{\hat{J}_z\}\cap\{{\rm
ker}\,\,
e_+\}$. This conjecture is strongly supported by the results of section
four, and those of \cite{tjin2}. We denote the representant of the full
cohomology by $\breve{J}_z=\hat{J}^{\rm HW}_z+ \cdots$
Combining this with eq. (\ref{Tgen}), we get that
\be
T= \mbox{e}^{\dis\ln\sqrt{C} e_0}\breve{J}_z \mbox{e}^{\dis - \ln\sqrt{C} e_0}.
\label{breve}
\ee

A priori, one can only say that the operator algebra of $\{T^{(j\a)}\}$ closes
modulo BRST exact terms. However, as  the cohomology was computed on a
reduced
complex which has no fields with negative ghost number, there are no BRST exact
terms at ghostnumber 0 in the reduced complex.

Computing T explicitly has obviously to be done on a case by case basis.
However the energy momentum tensor can be explicitly constructed. One easily
checks that
\bea
T_{\rm EM}&=&\frac{1}{x(k+\tilde{h})} str J_zJ_z -\frac{1}{8xy}str e_0 \del J_z
+ \sum_{j,\,\a_j}\frac{\k y}{j+\frac 1 2 }\t^{j\a}\del\t^{j\a}\nonu
&&-\sum_{j,\,\a_j}\frac{\k y}{j+\frac 1 2 }r^{j\a}\del\bar{r}^{j\a}
+\sum_{j,\,\a_j}\frac{\k y}{j+\frac 1 2 }\bar{r}^{j\a}\del r^{j\a}\nonu
&&+\frac{1}{4x}strb[e_0,\del c]-\frac {1}{2x}str b\del c+\frac{1}{4x}str
\del b
[e_0,c]
,\label{Timp}
\eea
where $e_0=t_{(10,0)}$,
is indeed BRST invariant. This current differs from the currents previously
studied by a BRST exact term:
\be
T_{\rm exact}=Q\left(-\frac{2}{x(\k+\tilde{h})}str b\hat{J}_z\right)
\ee
and we get:
\be
T^{(1,0)}=T+T_{\rm exact}=\frac{\k}{x(\k+\tilde{h})}str\left(
e_-+\t+r+\bar{r}\right)\hat{J}_z
+\mbox{more ...}\label{Tnimp}
\ee
{}From this we read off that the normalization constant $C$ is given by
\be
C=\frac{4y\k}{\sqrt{2}(\k+\tilde{h})}.
\ee
Eq. (\ref{Tnimp}) can easily be obtained from  eq. (\ref{Timp}) by
supplementing eq. (\ref{Timp}) with the constraints
\be
\P_-\left( \del g g^{-1} -e_- -r-\bar{r}-\t\right)=0.
\ee

We now analyze the various parts of the energy-momentum tensor, eq.
(\ref{Timp}). The fact that it differs by a BRST exact form from the ``true''
energy-momentum tensor does not change the value of the central charge. The
first term is the Sugawara tensor for $\hat{g}$ with central charge:
\be
c_{\rm Sug}=\frac{\k(d_B-d_F)}{\k+\tilde{h}}.\label{cccS}
\ee
The second term is the so-called improvement term. Given an affine current
$J\in\P_{-m}\bar{g}$, $m>0$, it takes care that the conformal dimension of $J$
is given by $-m+1$. Its contribution to the central charge is
\be
c_{\rm imp}=-6y\k.
\ee
The next term is the energy-momentum tensor for the auxiliary $\t$ fields with
central charge
\be
c_\t=\frac 1 2 \dim \left(\P^F_{1/2}\bar{g}\right),
\ee
and the two next terms give the energy-momentum tensor for the auxiliary $r$
and $\bar{r}$ fields with central charge
\be
c_{r\bar{r}}=-\frac 1 2 \dim \left(\P^B_{1/2}\bar{g}\right).
\ee
Finally, the last terms form the energy-momentum tensor for the ghost-antighost
system. To each generator $t$ of the gauge group, $t\in\P_m\bar{g}$ where
$m>0$, we associated a ghost $c\in\P_m\bar{g}$ with conformal dimension $m$ and
an anti-ghost $b\in\P_{-m}\bar{g}$ with conformal dimension $-m-1$. Such a pair
contributes $\mp (12 m^2-12m+2)$ where we have $-$ ($+$ resp.) if $b$ and $c$
are fermionic (bosonic resp.). As such the total contribution to the central
charge coming from the ghosts is given by
\be
c_{\rm ghost}=-\sum_{j,\a_j}(-)^{(\a_j)}2j(2j^2-1)+\frac{1}{2}\left( \dim
\left(\P^B_{1/2}\bar{g}\right) -\dim
\left(\P^F_{1/2}\bar{g}\right)\right)\label{cccg}
\ee
Adding all of this together, we obtain the full expression for the total
central charge $c$ as a function of the level $\k$:
\be
c=c_{\rm Sug}+c_{\rm imp}+c_\t+ c_{r\bar{r}}+c_{\rm ghost},\label{ccct}
\ee
where the individual contributions are given in eqs. (\ref{cccS}-\ref{cccg}).
Using the explicit form for the index of embedding $y$ given in eq.
(\ref{indexem}) and some elementary combinatorics, we can rewrite eq.
(\ref{ccct}) in the following, very recognizable form:
\be
c=\frac 1 2 c_{\rm crit} - \frac{(d_B-d_F)\tilde{h}}{\k+\tilde{h}} - 6 y (\k
+\tilde{h}),\label{cpretty}
\ee
where $c_{\rm crit}$ is the critical value of the central charge for the
extension of the Virasoro algebra under consideration:
\be
c_{\rm crit}=\sum_{j,\a_j}(-)^{(\a_j)}(12 j^2+12j+2).\label{ccrit}
\ee

We now turn to the effective action, which we determine along the same lines as
those followed in section 5.1. The effective action is given by:
\be
\exp -W[\check{T}]=\int [\d g g^{-1}][d\t][dr][d\bar{r}][d A_{\bz}]\left(
\mbox{Vol}\left( \P_+\bar{g} \right) \right)^{-1}\exp-\left( {\cal S}_1
+\frac{1}{4\p x y} \int str h\left( T-\check{T}\right)\right),\label{okok2}
\ee
where the sources $\check{T}$ are given by
\be
\check{T}\equiv\sum_{j,\a_j}\check{T}^{(j,\a_j)}t_{(jj,\a_j)}.
\ee

The road to follow is now precisely analogous to section 5.1. Choosing the
Drinfeld-Sokolov gauge, we obtain an explicit expression for the effective
action:
\be
W[\check{T}]=\k_c S_-[g]\label{endresult}
\ee
where
\be\k_c=\k+2\tilde{h}
\ee
and from eq. (\ref{cpretty}) we get the central extension as a function of the
level
\be
c=\frac 1 2 c_{\rm crit} - \frac{(d_B-d_F)\tilde{h}}{\k_c-\tilde{h}} - 6 y
(\k_c -\tilde{h}),\label{vv1}
\ee
or more usefully, the level as a function of the central charge:
\be
12 y \k_c=12 y\tilde{h}-\left(c-\frac 1 2 c_{\rm crit}\right)-
\sqrt{\left(c-\frac 1 2 c_{\rm crit}\right)^2- 24 (d_B-d_F)
\tilde{h}y}\label{vv2}
\ee
Eqs. (\ref{vv1}) or (\ref{vv2}) provide us with all-order expressions for the
coupling constant renormalization. We now turn to the wavefunction
renormalization.

The WZW model in eq. (\ref{endresult}) is constrained by
\be
\del g g^{-1}+\frac{\a_\k}{\k}\frac{1}{4xy}str \left\{\P_{\rm NA}\left(\del g
g^{-1}\right)\P_{\rm NA}\left(\del g g^{-1}\right)\right\}e_+ =
\frac{\k}{\a_\k} e_-+\frac{2}{\a_\k}\mbox{e}^{\dis -\ln\sqrt{C} e_0}\check{T}
\mbox{e}^{\dis\ln\sqrt{C} e_0},
\ee
where  $\P_{\rm NA}\bar{g}$ are those elements of $\P_0 \bar{g}$ which do not
belong to the Cartan subalgebra of $\bar{g}$, {\it i.e.} the centralizer of
$\slt$ in $\bar{g}$.
Performing a global group transformation
\be
g\rightarrow \mbox{e}^{\dis \ln \left(\sqrt{\frac{\a_\k}{\k}}\,\right) e_0 }g.
\ee
we bring the constraints in the standard form eq. (\ref{aastd}):
\be
\del g g^{-1}+\frac{1}{4xy}str \left\{\P_{\rm NA}\left(\del g
g^{-1}\right))\P_{\rm NA}\left(\del g
g^{-1}\right)\right\}e_+=e_-+\sum_{j,\a_j}
\frac{2\k^j}{C^j\a_\k^{j+1}}\check{T}^{(j\a_j)} t_{(jj,\a_j)}.
\ee
All computations were done using operator methods, so again $\a_\k$ is given by
$\a_\k=\k+\tilde{h}$. With this choice we get the following final expression
for the constraints:
\be
\del g g^{-1}+\frac{1}{4xy}str \left\{\P_{\rm NA}\left(\del g
g^{-1}\right))\P_{\rm NA}\left(\del g g^{-1}\right)\right\}e_+=e_-+\frac{1}{\k
+\tilde{h}} \sum_{j,\a_j}
\frac{1}{2^{\frac 3 2 j-1}y^{j}}
\check{T}^{(j\a_j)} t_{(jj,\a_j)},
\ee
which gives us, for the chosen normalization, eq. (\ref{Tgen}), of the
conformal currents, the wavefunction renormalization to all orders.

{}From eq. (\ref{vv2}), we deduce that for generic values of $\k$,
no renormalization of the coupling constant beyond one loop occurs if and only
if either $d_B=d_F$ or $\tilde{h}=0$ (or both). Both cases are only possible
for superalgebras. We get $d_B=d_F$ for $su(m\pm 1|m)$, $osp(m|m)$ and
$osp(m+1|m)$. The quadratic Casimir  in the adjoint representation
vanishes, {\it i.e.} $\tilde{h}=0$, for $su(m|m)$, $osp(m+2|m)$ and
$D(2,1,\alpha)$. Note that $P(m)$ and $Q(m)$ have not been considered,
since the absence of an invariant metric implies that no WZW models exist
for them. The nonrenormalization of the couplings is reminiscent of
nonrenormalization theorems \cite{nonrenor} for extended supersymmetry.
These imply that under suitable circumstances at most one loop corrections
to the coupling constants are present (the wave function renormalization
may have higher order contributions). Comparing our list with the
tabulation \cite{Sofra} of super-$W$ algebras obtained from a (classical)
reduction
of superalgebras, we find that many of them, though not all, have $N=2$
supersymmetry. For instance, there is an $sl_2$ embedding in $osp(3|2)$
which gives the super-$W_2$ algebra of \cite{josestany}, which contains
four fields (dimensions 5/2,2,2,3/2) and no $N=2$. Also, it seems that
all superalgebras based on the reduction of the unitary superalgebras
$su(m|n)$ contain an $N=2$ subalgebra, whereas our list contains only the
series $|m-n|\leq 1$. Clearly, the structural reason behind the lack of
renormalization beyond one loop remains to be clarified \cite{alexalex}.


\section{Examples}
\setcounter{footnote}{0}

As an application of the general framework developed in previous section, we
briefly study a few examples.

\subsection{Other $N=4$ Supergravities}
\setcounter{equation}{0}

The $N=4$ algebra given in eq. (\ref{alg2}) is only a special case of a one
parameter family of N=4 algebras. This one-parameter family of $N=4$
superconformal algebras occurs both in a linearized version, discovered in
\cite{n4}, generalizing the Ademollo et al. \cite{adem} $N=4$ algebra and a
non-linearly generated version, generalizing the Knizhnik-Bershadsky $SO(4)$
\cite{osp1,osp2} algebra, discovered in \cite{goddard}.

The latter differs essentially from the $N=4$ algebra given in eq. (\ref{alg2})
by the $GG$ OPE  for which at the right hand side, the $so(4)$ current algebra
gets broken to two commuting $su(2)$ current algebras, the relative strength of
which is determined by a parameter $\a\equiv k_+/k_-$, where $k_\pm$ are the
levels of the two $su(2)$ current algebras.
The central charge $c$ of the superconformal algebra is given by
\be
c=\frac{6k_+k_-}{k_++k_-}-3.\label{cnna}
\ee
For $k_+=k_-$ the superconformal algebra is isomorphic to eq. (\ref{alg2}) for
$N=4$. The subalgebra of transformations globally defined on the sphere is
on-shell isomorphic to $D(2,1,\a)$. The resulting effective supergravity theory
is given by a constrained $D(2,1,\a)$ WZW model. The  methods and results are
very similar to those used and obtained for $osp(4|2)$ in sections 4 and 5.
E.g. the level as function of the central charge is precisely given by eq.
(\ref{renormc}) for $N=4$.

Again the superconformal algebra can be linearized by adding 4 free fermions
and a $U(1)$ current. Using the results of \cite{kris1}, we find that the level
of the $D(2,1,\a)$ current algebra is related to the central extension $c_{\rm
lin}=c+3$ by
\be
\k_c=-2c_{\rm lin}.
\ee

Finally, there is one more ``standard'' $N=4$ superconformal algebra. It can be
obtained by folding or twisting \cite{n4} the previously mentioned {\it linear}
$N=4$ algebras. The resulting superconformal algebra has besides the
energy-momentum tensor, 4 dimension 3/2 supercurrents and an $so(3)$ affine Lie
algebra. After the twist the algebra acquires a central charge $\tilde{c}$:
\be
\tilde{c}=6k_+
\ee
and the level of the $so(3)$ current algebra is given by $2k_+$ (the factor 2
explains why we called this an $so(3)$ algebra).

The subalgebra of transformations globally defined on the sphere is isomorphic
to $su(1,1|2)$. As such we expect that the corresponding effective supergravity
theory is given by a constrained $su(1,1|2)$ WZW model. For the principal
embedding of $\slt$ in $su(1,1|2)$, one finds that the adjoint representation
of $su(1,1|2)$ branches to a $j=1$, 4 $j=1/2$ and 3 $j=0$ (generating an
$so(3)$ subalgebra) irreducible representations of $\slt$. The critical
dimension is $c_{\rm crit}=-12$ and due to the nature of the embedding we find
$y=1$ (Formula (\ref{indexem}) can not be used, but the value is obvious).
Considering the principal embedding of $\slt$ in
$su(1,1|m)$ where one finds, using eq. (\ref{indexem}) that $y=1$ for every
$m$. Taking all of this together, we find
\be
-6\k_c=\tilde{c}+1.
\ee
The factor 1 at the rhs combined with the results of \cite{kris1}, suggest that
adding a $U(1)$ current to the superconformal algebra, will again yield a
theory where the coupling constant does not get renormalized. One such a theory
which comes to mind is the one where we realize the $N=4$ superconformal
algebra in terms of 2 complex fermions and 4 ``symplectic'' bosons
\cite{golive}. For this particular theory one gets $\k_c=0$. As such, this
theory will only have a finite number of degrees of freedom. A further study of
this extended topological gravity would be interesting.

\subsection{$W$ Gravity}

The $WA$ algebras, which are ususally called $W$ algebras, are the most studied
non-linearly generated conformal algebras. They are characterized by the fact
that the subalgebra of transformations, globally defined on the sphere, forms
on-shell, {\it i.e.} putting the non-linear terms to zero, an $\sln$ algebra.
For given $n$, the different $W$ algebras are classified by the inequivalent,
non-trivial embeddings of $\slt$ in $\sln$. And the procedure of section six
can be applied to construct both the induced action, {\it i.e.} essentially
realize the $WA$ algebra in terms of a gauged $\sln$ WZW model, and the
effective action.

It is known \cite{dynkin} that the inequivalent embeddings of $\slt$ in $\sln$
are completely characterized by the branching of the fundamental representation
of $\sln$, $\unn$ in irreducible representations of $\slt$\footnote{In other
words, they are in one to one correspondence with the partitions of $n$.}:
\be
\unn=\bigoplus_{j\in\frac 1 2 {\bf N}}n_j\cdot[2j+1].\label{sle1}
\ee
As we only consider non-trivial embeddings, we have to supplement eq.
(\ref{sle1}) with the condition:
\be
q\equiv\sum_{j\in\frac 1 2 {\bf N}}n_j<n.\label{sle2}
\ee
The ``standard'' $W$ algebras, correspond to the partition for which the only
nonvanishing $n_j$ is $n_{(n-1)/2}=1$.

{}From now on, we work in a given embedding. The branching of the adjoint
representation of $\sln$, $\underline{n^2-1}$, into irreducible representations
of $\slt$ follows immediately from $\unn
\otimes\overline{\unn}=\underline{1}\oplus \underline{n^2-1}$. We get
\bea
\underline{1}\oplus \underline{n^2-1}&=&
\bigoplus_j n_j^2\cdot \bigoplus_{J=0}^{2j}[2J+1]\oplus\nonu
&&\bigoplus_{j>j'}2 n_jn_{j'}\cdot \bigoplus_{J=j-j'}^{j+j'}[2J+1].\label{sle3}
\eea
{}From eq. (\ref{sle3}), we can read the field content of the corresponding
$WA_n$ algebra. The algebra has $qn-1-2\sum_{j>j'}n_jn_{j'}(j-j')$ currents,
with conformal dimensions given by $J+1$, where $J$ labels the $\slt$
representations in the branching of
$\underline{n^2-1}$. The affine subalgebra of the conformal algebra, is
determined by the centralizer $\P_{\rm NA}\sln$, of $\slt$ in $\sln$:
\be
\P_{\rm NA}\sln =\bigoplus_{j\in\frac 1 2 {\bf N}}sl(n_j,{\bf R})\oplus
(r-1)\cdot u(1),
\label{sle4}
\ee
where $r$ is the number of different values of $j$ for which $n_j\neq 0$.
Combining eq. (\ref{sle1}) with the well-known action of $\slt$ on irreducible
representations:
\bea
e_0|jm>&=&2m|jm>\nonu
e_\pm|jm>&=&\sqrt{(j\mp m )(j\pm m+ 1)}|j m\pm 1>,\label{sle5}
\eea
we immediately obtain the explicit expression of the $\slt$ generators in terms
of $n\times n$ matrices.

The only two quantities which remain to be computed is the index of embedding
$y$ and the critical dimension $c_{\rm crit}$.
The index of embedding, eq. (\ref{indexem}), is very easily computed using the
$sl(2)$ characters $\c_J(\q )=\sin ((2J+1)\q)/\sin(\q)$ and using the fact that
\be
\lim_{\q\rightarrow 0}\frac{\del^2 \c_J(\q )}{\del^2\q}=-\frac 4 3
J(J+1)(2J+1).\label{sle6}
\ee
One finds the following, very simple expression:
\be
y=\frac 2 3 \sum_{j\in\frac 1 2 {\bf N}}n_jj(j+1)(2j+1).
\ee
The critical dimension is obtained through direct computation, {\it i.e.}
combining eqs. (\ref{sle3}) and (\ref{ccrit}):
\be
c_{\rm crit}=6qy+2qn-2+12n\sum_jn_j
j(j+1)+4\sum_{j>j'}n_jn_{j'}(j-j')(1-2(j-j')^2).
\ee
Using this, $\tilde{h}=n$, $d_B=n^2-1$ and $d_F=0$, one finds from eq.
(\ref{vv2}) the level as a function of the central charge.

All these ingredients combined fully determine the effective action in terms of
a constrained $\sln$ WZW model.

Comparing these results to some of the known ones, like $W^{(1)}_3$ gravity in
\cite{ssvprep,ssvnc,dbg} and $W^{(2)}_32$ gravity in \cite{mic,tjin2}, we
find complete agreement.

\subsection{$N=2$ $W_n$ Gravity}

The principal embedding of $\slt$ in $su(n|n-1)$ yields $N=2$ $W_n$ algebras.
Twisted versions of these systems are very relevant in the study of
non-supersymmetric $W_n$ strings \cite{wolf}. $\slt$ gets embedded in the
$su(n)$ subalgebra such that the $\unn$ of $su(n)$ branches to the $[n]$ of
$\slt$. The adjoint of $su(n|n-1)$ branches to $n$ ``$N=2$ multiplets'':
$(j,j+1/2,j+1/2,j+1)$ where $j\in \{0,2,\cdots , n-2\}$:
\be
\mbox{adjoint}(su(n|n-1))=\bigoplus_{j=0}^{n-2}\left\{[2j+1]+2\cdot
[2j+2]+[2j+4] \right\}.
\ee
In a $(j,j+1/2,j+1/2,j+1)$ multiplet, the first irreducible representation $j$
arises from the bosonic $su(n-1)+u(1)$ factor of $su(n|n-1)$, while the two
$j+1/2$ irreducible representations are fermionic. The $j+1$ irreducible
representation comes from the  $su(n)$ subalgebra. These superconformal
algebras contain as well the $W_n$ algebra as a
subalgebra as the N=2 Virasoro algebra and as such deserve the name of $N=2$
$W_n$ algebras.

The critical dimension is easily computed and gives $c_{\rm crit}= 6 n-6$.
The index of embedding was computed in previous section and gives
\be
y=\frac 1 6 n(n-1)(n+1).
\ee
Combining this we find
\be
c=3(n-1)-n(n-1)(n+1)(\k_c-1).
\ee

In section 6 we classified all superalgebras which give rise to extensions of
$d=2$ gravity, where no coupling constant renormalization beyond one loop
occurs. The example we just analyzed is a member of this class. These theories
do indeed have an $N=2$ supersymmetry.

\section{Conclusions}
\setcounter{equation}{0}
\setcounter{footnote}{0}

Inequivalent, nontrivial embeddings of $\slt$ in a (super) Lie algebra are
associated to extensions of the Virasoro algebra. These extended conformal
algebras were realized in section 6 by considering a WZW model in which a
chiral, solvable group is gauged. This description forms a perfect starting
point to study the associated extension of $d=2$ gravity in the chiral gauge.
These extensions assume various forms: higher spin gauge fields, more
gravitons, fermionic fields, Yang-Mills type (super)symmetries, ... The
description in terms of a gauged WZW model allowed us to obtain an all order
expression for the effective action. The effective action turned out to be a
constrained WZW model and we gave all-order expressions for the coupling
constant renormalization and the wavefunction renormalization.

In sections 2-5 we presented a detailed study of $SO(N)$ supergravity. The
cases $N=2$, $N=3$ and $N=4$ have the particular feature that the coupling
constant does not get renormalized beyond one loop.
If one considers $N=4$ supergravity based on the linear $N=4$ superconformal
algebra, one finds that no renormalization at all occurs!

This is consistent with the non-renormalization theorems
\cite{nonrenor}for extended
supersymmetry. While these theorems do not say anything on
the wavefunction
renormalization, they predict at most a one loop renormalization for the
coupling constant.  The one loop contribution is basically due to the infinite
tower of ghosts which arises when one expresses the constrained superfields in
terms of unconstrained ones ({\it i.e.} solving the constraints). In the case
of $N=4$, it can occur that even the one loop contribution vanishes. This
clearly occurs here. A detailed study of the non-renormalization effects
requires a superspace formulation of these theories.

The $SO(2)$, $SO(3)$ and $SO(4)$ supergravity models are not isolated cases. In
section 6 we derived an all-order expression for the coupling constant
renormalization, eqs. (\ref{vv1},\ref{vv2}). From this we obtained an
exhaustive list of all the models where this phenomenon occurs. Not all of
these models posses an $N=2$ supersymmetry as part of the total symmetry. Some
of them are characterized by a super-$W_2$ structure. This superconformal
algebra has currents of dimension $3/2$, $2$, $2$ and $5/2$. Though we do
recognize some kind of $N=2$ supermultiplet here, the algebra is definitely not
the $N=2$ algebra. A further study of these structures is underway.

Let us now adress the question of non-critical strings. Given an embedding of
$\slt$ in $\bar{g}$, we consider the corresponding $(p,q)$ minimal model as the
matter sector of the string theory. Its central charge $c_M$ is given by eq.
(\ref{cpretty}), where $\k_{M}+\tilde{h}=p/q$. In order to cancel the conformal
anomaly, we need to supplement the matter sector by a gauge sector whose
central charge $c_L$ is again given by eq, (\ref{cpretty}) but now
$\k_{L}+\tilde{h}=-p/q$. The corresponding $W$ string is now determined by
currents
\be
T_{\rm tot}=T_{M} +  T_{L}
\ee
where $T_M$ and $T_L$ are of the form given in eq. (\ref{breve}) and a BRST
charge of the form
\be
Q=\frac{1}{2\p i}\oint str c(T_{\rm tot} +\frac 1 2 T_{ghost}),
\ee
where the ghost system contributes $-c_{\rm crit}$ to the central charge. Of
course the whole problem is to construct the ghost system.

Ultimately, the most straightforward way to construct $W$ strings is departing
from a matter sector covariantly coupled to the gravitation theory. The
matter-sector is most elegantly described by a gauged WZW model in the
``conformal'' gauge, as in this gauge it has the extended conformal symmetry
for both the left and the right movers. In order to couple the matter action
covariantly to the extended gravity, one can again use WZW like techniques (in
some sense a generalization of the results of \cite{hv,dbgcov}). In a future
publication we will present a detailed analysis of such coupled, gauged WZW
models.

One issue to be resolved is the rigorous  proof of the conjecture on the
computation of the reduced cohomology in section 6. Once this is done, one can
study the question whether in the presence of auxiliary fields the Feigin-Fuchs
type free field realization of $T$ is, as in the purely bosonic case, obtained
by putting all fields but $\P_0\hat{J}_z$, $\t$, $\t$ and $\t$ to zero in eq.
(\ref{breve}).

A most challenging problem is the understanding of the geometry behind the
extensions of $d=2$ gravity. By geometry, we mean something very simple.
The Virasoro algebra appears as the algebra of residual symmetry after
gauge fixing a theory invariant under general coordinate transformations in
$d=2$. The question is whether a similar statement can be made for extensions
of $d=2$ gravity. In particular, the geometric significance of the
non-linearities in the extensions of the Virasoro algebra remain to be
understood. However, the $N=3$ and $N=4$ supersymmetric theories might provide
clues for the solution of this problem. As we mentioned earlier, the
non-linearly generated $N=3$ and $N=4$ algebras can be linearized by adding
free fields to the system. In the linear case, as both $N=3$ and $N=4$
superspace have been constructed, the geometry is well understood. The relation
between the linear and non-linear algebras might enable one to learn something
about the geometry of the non-linear algebras. Finally, a most exciting
application of the methods developed in this paper would be the study of
reductions of continuum algebras \cite{saveliev} which would presumably lead to
integrable theories in $d>2$! Work in these directions is in progress.

\vspace{1.5cm}

\noindent
{\bf Acknowledgements}: We would like to thank Orlando Alvarez, Jan De Boer,
Alex Deckmyn, Wolfgang Lerche, Micha Savelev, Kareljan Schoutens, Wati Taylor,
Jean Thierry-Mieg, Peter van Nieuwenhuizen,  and especially Martin Ro\v{c}ek
for many illuminating discussions.


\setcounter{section}{0}
\setcounter{equation}{0}
\startappendix
\section{Wess-Zumino-Witten Models}
\setcounter{footnote}{0}

In this appendix we give some essential properties of WZW models.
We start by summarizing some properties of super Lie algebras. Given a super
Lie algebra with generators $\{ t_a ; a\in\{ 1,\cdots,d_B+d_F\}\}$, where
$d_B$ ($d_F$) is the number of bosonic (fermionic) generators, we denote the
(anti)commutation relations by
\be
[t_a, t_b]=t_at_b-(-)^{(a)(b)}t_b t_a=f_{ab}{}^c t_c,
\ee
where for $t_a$, $(a)=0\ (1)$ when $t_a$ is bosonic (fermionic).
We adopt the convention that $[A,B]$ stands for the anticommutator if both $A$
and $B$ are fermionic, else it is a commutator. We also take
$X t_a= (-)^{(X)(a)} t_a X $ where $X$ is not Lie algebra valued.
{}From the Jacobi identities one shows that the adjoint representation is given
by
\be
[t_a]_b{}^c\equiv f_{ba}{}^c.
\ee
The Killing metric $g_{ab}$ is defined by
\be
f_{ca}{}^d
f_{db}{}^c(-)^{(c)} = - \tilde{h} g_{ab},
\ee
where $\tilde{h}$ is the dual Coxeter
number. Though this is perfect for ordinary Lie algebras, this is not
sufficient for super algebras as the dual Coxeter
number might vanish in this case\footnote{The possibility of a vanishing
quadratic Casimir in the adjoint representation, has the interesting
consequence that in that case,  the affine Sugawara construction always gives
a value for $c$ which is independent of the level of the affine super Lie
algebra. E.g. for $D(2,1,\a)$, one always has $c=1$.}. More generally we have
\be
str(t_at_b)\equiv [t_a]_{\a}{}^{\b}[t_b]_{\b}{}^{\a}(-)^{(\a )}\equiv -x g_{ab}
\ee
where $x$ is the index of the representation. Obviously we have $x=\tilde{h}$
in the adjoint representation. A contraction runs from upper left to lower
right, {\it e.g.} $A^aB_a$. The Killing metric is used to raise and lower
indices according to this convention (implying $g^{ac}g_{bc}=\d^a_b$):
\be
A^a=g^{ab}A_b\qquad A_a=A^bg_{ba}.
\ee
We tabulate some properties of the (super) Lie algebras which
appear in this paper:

\begin{center}
\begin{tabular}{||c||c|c|c|c|c||}  \hline
 & & & & & \\
algebra    & bosonic & {$d_B$}    & {$d_F$} & {$\tilde{h}$} & {$x_{fun}$}   \\
& subalgebra& && &\\ \hline
{$sl(n)$}  & {$sl(n)$}  &{${\scriptstyle n^2-1}$}  &{${\scriptstyle  0}$}
& {${\scriptstyle n} $}         & {${\scriptstyle \frac 1 2} $} \\ \hline
{$so(n)$}  & {$so(n)$}  &{$ {\scriptstyle \frac 1 2 n(n-1)}$} &{$ {\scriptstyle
 0}$} & {${\scriptstyle n-2 }$}   & {$ {\scriptstyle  1 }$}            \\
\hline
{$osp(n|2)$}&{$sl(2)+so(n)$}& {$ {\scriptstyle \frac 1 2 (n^2-n+6) }$}&{$
{\scriptstyle 2n }$} & {${\scriptstyle \frac 1 2 (4 - n) }$} &{$ {\scriptstyle
\frac 1 2 } $}\\ \hline
{$D(2,1,\a)$}&{$sl(2)+su(2)+su(2)$}&{${\scriptstyle 9}$} &{$ {\scriptstyle 8}
$} & {$ {\scriptstyle 0}$} & {${\scriptstyle -}$} \\ \hline
{$su(1,1|2)$}&{$sl(2)+su(2)$}&{${\scriptstyle 6}$} &{$ {\scriptstyle 8} $} & {$
{\scriptstyle 0}$} & {${\scriptstyle \frac 1 2 }$} \\ \hline
{$su(m|n)$}&{$su(m)+su(n)+u(1)$}&{${\scriptstyle m^2+n^2-1}$} &{$ {\scriptstyle
2mn} $} & {$ {\scriptstyle m-n}$} & {${\scriptstyle \frac 1 2 }$} \\
{$m\neq n$}& & & & & \\ \hline
\end{tabular}
\end{center}

\noindent
We denoted by $x_{fun}$, the index of the fundamental (defining)
representation. For $D(2,1,\a)$, the size of the fundamental representation
depends on $\a$. The smallest representation which exists for all values of
$\a$ is the adjoint representation.

The WZW action $\k S^+[g]$ is given by
\be
\k S^+ [g]=  \frac{\k}{4\p x} \int d^2 x \;
str \left\{ \del g^{-1} \bar{\del} g \right\}
+ \frac{\k}{12\p x} \int d^3 x\; \e^{\a\b\g} \,
str \left\{ g_{,\a} g^{-1} g_{,\b}
g^{-1} g_{,\g} g^{-1} \right\}.
\label{nineteen}
\ee
It satisfies the Polyakov-Wiegman identity \cite{polwi},
\be
S^+[hg]=S^+[h]+S^+[g]-\frac{1}{2\pi x}\int str\Bigl( h^{-1}\del h \bdel g
g^{-1} \Bigr),\label{pwfor}
\ee
which is obtained through direct computation.
We also introduce a functional $S^-[g]$ which is defined by
\be
S^-[g]=S^+[g^{-1}].
\ee
The equations of motion follow from
\bea
\d  S^+[g]&=&\frac{1}{2\p x}\int str\left\{\bdel(g^{-1}\del g)g^{-1}\d
g\right\}\nonu
&=&\frac{1}{2\p x}\int str\left\{\del(\bdel g g^{-1})\d g g^{-1}\right\},
\eea
which is solved by putting $g\equiv g(\bz )g(z)$ where $\del g(\bz )=\bdel g(z
)=0$.
As such we get that the currents
\bea
J_z&=&-\frac{\k}{2}g^{-1}\del g \nonu
J_{\bz}&=&\frac{\k}{2}\bdel g g^{-1}
\eea
are conserved. This implies the affine symmetries
\bea
\d J_z^a&=&-\frac{\k}{2} \del\h^a- (-)^{(b)(c)}f_{bc}{}^{a}\h^bJ_z^c\nonu
\d J_{\bz}^a&=& \frac{\k}{2}\bdel\bar{\h}^a+ (-)^{(b)(c)}f_{bc}{}^{a}
\bar{\h}^b J_{\bz}^c
\eea
where
\be
\bdel\h^a=\del\bar{\h}^a=0.
\ee
{}From
\be
\d J_z^a(x)=\frac{1}{2\p i}\oint_x dy J_z^b(y)\h_b(y)J^a_z(x),
\ee
we get the OPE of an  affine Lie algebra of level $\k$:
\be
J^a_{z} (x) J^b_{z} (y) = - \frac{\k}{2} g^{ab} (x-y)^{-2} + (x-y)^{-1}
(-)^{(c)}f^{ab}{}_c
J^c_{z} (y) + \cdots,
\label{ven}
\ee
and similarly for $J_{\bz}$. The Sugawara tensor is given by
\be
T=\frac{1}{x\left(\k+\tilde{h} \right)}str J_zJ_z,
\ee
and it satisfies the Virasoro algebra with the central extension given by:
\be
c=\frac{k(d_B-d_F)}{k+\tilde{h}}.
\ee

As an example we list the OPEs for $\osp$, which will be used extensively
throughout  section four.

\bea
J^0(x)J^0(y)&=&\frac \k 8 (x-y)^{-2}\nonu
J^{\pp}(x)J^=(y)&=&\frac \k 4 (x-y)^{-2}+(x-y)^{-1}J^0(y)\nonu
J^0(x)J^{\stackrel{\scriptstyle\!\! \pp}{=}}(y)&=&\pm \frac 1 2 (x-y)^{-1}
J^{\stackrel{\scriptstyle\!\!\pp}{=}}(y)\nonu
J^0(x)J^{\pm a}(y)&=&\pm \frac 1 4 (x-y)^{-1}J^{\pm a}(y)\nonu
J^{\stackrel{\scriptstyle\!\!\pp}{=}}(x)J^{\mp a}(y)&=&\frac 1 2 (x-y)^{-1}
J^{\pm a}(y)\nonu
J^{i}(x)J^{j}(y)&=&\frac \k 8 \d^{ij}(x-y)^{-2}-\frac{ \sqrt{2}}{4}(x-y)^{-1}
f_{ij}{}^kJ^k(y)\nonu
J^{i}(x)J^{\pm a}(y)&=&\frac{\sqrt{2}}{4}(x-y)^{-1}\l_{ab}{}^iJ^{\pm b}\nonu
J^{+a}(x)J^{-b}(y)&=&\frac \k 8 \d^{ab}(x-y)^{-2}+\frac 1 4
\d^{ab}(x-y)^{-1}J^0(y)+\frac{\sqrt{2}}{4}(x-y)^{-1}\l_{ab}^iJ^i(y)\nonu
J^{\pm a}(x)J^{\pm b}(y)&=&\mp \frac 1 4 \d^{ab}(x-y)^{-1}
J^{\stackrel{\scriptstyle\!\!\pp}{=}}(y),\label{ospope}
\eea
where the metric we used is given by
\be
g_{\pp \, =}=-2,\qquad g_{00}=-4,\qquad g_{+a-b}=-4\d_{ab},\qquad
g_{ij}=-4\d_{ij}.
\ee
Note that the $J_iJ_j$ OPE can be rewritten as
$J_i(x)J_j(y)=2\k\d_{ij}(x-y)^{-2}+\sqrt{2}(x-y)^{-1}f_{ij}{}^kJ_k$, from which
we observe, as was to be expected, that the $SO(N)$ level is even and negative.

\setcounter{equation}{0}

\startappendix
\section{Induced Gauge Theories}

In this appendix we review some basic and well-known results on gauged WZW
models \cite{wzw,polwi,orlando,dive}.

Consider the induced action, $\G[A_{\bz}]$, for the gauge fields $A_{\bz}$,
\be
e^{\dis -\G[A_{\bz}]} = \, \langle
  \exp \, - \frac{1}{\p x} \int d^2 x \;
  str \left\{ J_z(x) A_{\bz}(x) \right\} \rangle ,
\label{twelve}
\ee
where $J_z$ satisfies eq. (\ref{ven}).
The gauge transformations
\be
\d A_{\bz}=\bar{\del} \h + [\h, A_{\bz}]\ ,
\label{gf}
\ee
are anomalous
\be
\d \G [A_{\bz}] = - \frac{\k}{2\p x} \int d^2 x \, str \left\{ \h \del A_{\bz}
\right\}.
\label{thirteen}
\ee
Defining
\be
u^a_z(x)= - \frac{2\p}{\k}g^{ab} \frac{\d \G[A_{\bz}]}{\d A_{\bz}^b(x)},
\label{sixteen}
\ee
we deduce from eqs. (\ref{thirteen}) and (\ref{gf}) the following
Ward identity
\be
\bar{\del}  u_z - [A_{\bz}, u_z ] =\del A_{\bz} .
\label{fifteen}
\ee
The Ward identity is independent of $\k$, therefore
\be
\G [A_{\bz}]=\k\G^{(0)}[A_{\bz}],
\ee
where $\G^{(0)}[A_{\bz}]$ is independent of $\k$.
In \cite{polwi,orlando}, it was observed  that eq.\ (\ref{fifteen}) states that
the curvature for the Yang-Mills fields $\{A,u\}$ vanishes. This condition
is solved by parametrizing $A_{\bz}$ as $A_{\bz}\equiv \bar{\del} g g^{-1}$ and
$u_z$ as
$u_z\equiv\del g g^{-1}$. Introducing the WZW functional $S^+[g]$,
we easily find that $\G^{(0)} [A_{\bz}]$ is given by
\be
\G^{(0)} [A_{\bz}=\bdel g g^{-1}]=-S^+[g].
\ee

We could as well have performed the previous analysis starting from an
anti-holomorphic affine Lie algebra. The induced action is then given by
\be
e^{\dis -\overline{\G}[A_z]} = \, \langle
  \exp \, - \frac{1}{\p x} \int d^2 x \;
  str \left\{ J_{\bz}(x) A_z(x) \right\} \rangle ,
\label{atwelve}
\ee
where
\be
\overline{\G}[A_z=\del g g^{-1}]= -\k S^+[g^{-1}] \equiv -\k S^-[g].
\ee

We now consider the generating functional of connected Greens functions with
propagating $A_{\bz}$ fields, $W[u_z]$\footnote{The source $u_z$ is obviously
different from $u_z$ defined in eq. (\ref{sixteen}). We hope that this does not
cause any confusion.} :
\be
e^{\dis -W[u_z]}=\int [dA_{\bz}] e^{\dis -\G[A_{\bz}]+\frac{1}{2\p x}
  \int  \; str (u_zA_{\bz}) }\label{wu}
\ee
The Legendre transform of $\G^{(0)}[A_{\bz}]$,
\be
W^{(0)}[u_z]=\min_{\{ A_{\bz}\} }\left( \G^{(0)}[A_{\bz}]-\frac{1}{2\pi x}\int
str \left( u_z\,A_{\bz}\right)\right)
\ee
is  explicitly given by
\be
W^{(0)}[u_z\equiv \del g g^{-1}]=S^-[g].\label{qquan1}
\ee
It can be argued in several different ways that $W(u_z)$ is given by
\be
W(u_z)=\hat{\k}W^{(0)}[Z_{\k}u_z].
\ee
The different computations of the renormalised coupling $\hat \k$
agree with each other, but the values of the current renormalisation
factor $Z_{\k}$ differ. First we present two different arguments leading
to the value used in the text. Then, for completeness, we also sketch
two other lines of reasoning.

In eq.(\ref{wu}), we parametrise the integration variable by $A_{\bz}=
\bdel g g^{-1}$. For the Jacobian, we use
\be
[dA_{\bz}]=[dgg^{-1}]\det \overline{D}[A_{\bz}\equiv\bdel g g^{-1}]=[dgg^{-1}]
\exp2\tilde{h}S^+[g].\label{yoyo}
\ee
and obtain
\be
e^{\dis -W[u_z]}=\int [dg g^{-1}] e^{\dis (\k+2\tilde{h})S^+[g] +\frac{1}{2\p
x} \int  \; str (u_z\bdel g g^{-1}) }.
\label{pathint}
\ee
Since the currents $\bdel g g^{-1}$ form an antiholomorphic affine
Lie algebra with level $K=-(\k+2\tilde{h})$,
$W[u_z]$ is the corresponding induced action, cfr. eq. (\ref{atwelve}).
The only remaining problem is to identify the proportionality constant in
$\bdel g g^{-1}\propto J_{\bz}$. Given a current algebra of level $K$, we put
\be
J_{\bz}=\frac{\a_K}{2}\bdel g g^{-1}.\label{defalfa}
\ee
So we conclude that
\be
W[u_z]=(\k+2\tilde{h})W^{(0)}\left[ -\frac{u_z}{\a_{-\k-2\tilde{h}}} \right]
\ee
and we identify
\bea
\hat{\k}&=&\k+2\tilde{h}\nonu
Z_\k&=&-\frac{1}{\a_{-\k-2\tilde{h}}}.
\eea
Using OPE techniques, it is argued in \cite{kz} that for a
current algebra of level $K$, the conventionally normalised currents are
$J_{\bz}=\frac{K+\tilde{h}}{2}\bdel g g^{-1} \, $. This follows from
consistency requirements in the operator product algebra
of the currents with $g$. In our case we have
that $K=-(\k+2\tilde{h})$, so accepting this argument we find
\be
\a_\k=-\k-\tilde{h}\label{kzalfa}
\ee
and
\be
Z_\k=\frac{1}{ \k+\tilde{h}}.\label{qquan3}
\ee
This is the value used in the main text. It is compatible with the value
found in perturbation theory using the method described in section 5.2.
This conclusion follows immediately from eq.(\ref{Wusemicl}), so the
reasoning will not be repeated here.

A different argument rests on the invariance of the Haar measure and the
Polyakov-Wiegmann formula. Parametrizing $u_z$ as
\be
u_z\equiv (\k+2\tilde{h})\del h h^{-1}
\ee
we find, using eq.(\ref{pwfor}), that eq.(\ref{pathint}) becomes:
\be
e^{\dis -W[u_z]}=e^{\dis -(\k+2\tilde{h})S^-[h]}
\int [dg g^{-1}] e^{\dis (\k+2\tilde{h})S^+[h^{-1}g]}.
\ee
Assuming that we use a regulator which leaves the Haar measure invariant, we
can drop the functional integral and we find that
\be
Z_\k=\frac{1}{ \k+2\tilde{h}}.\label{value1}
\ee
This corresponds to the classical value for $\a_K$: $\a_K=K$.

It is not difficult to reproduce this value by setting up the
semiclassical computation differently.
In fact, if in the perturbative calculation we factorize the
determinants as follows:
\be
\det\left(\frac{ D[u_z/\k]}{\overline{D} [A^{\rm cl}_{\bz}(u_z)]}
\right)=\frac{\det D[u_z/\k]\overline{D} [A^{\rm cl}_{\bz}(u_z)]}{\left(\det
\overline{D} [A^{\rm cl}_{\bz}(u_z)]\right)^2},
\ee
and we compute the determinants in the numerator with a vector gauge
invariant regulator, we find eq. (\ref{value1}) back.

This leads us to yet another method to compute $W[u_z]$:
the KPZ approach, \cite{kpz}. The covariant induced action (which can be
obtained from the computation of {\it e.g.} $\det \left(\overline{D}[A_{\bz}]
D[A_z]\right)$ with a gauge invariant Pauli-Villars regulator) is given by
\be
\G [A_{\bz}]+\overline{\G}[A_z]-\frac{\k}{2\pi x}\int
str\left\{A_{\bz}\,A_z\right\},\label{covindaction}
\ee
where $A_{\bz}$ transforms as in eq. (\ref{gf}) and $ \d A_z =\del \h + [\h,
A_z]$. Now we want to use (\ref{covindaction}) as a quantum action.
Using the gauge freedom to fix
$A_z\equiv \hat{A_z}$ yields the gauge fixed, BRST-invariant action:
\be
\G[A_{\bz},\hat{A_z},b,c]=\G [A_{\bz}]+\overline{\G}[\hat{A_z}]- \frac{\k}{2\pi
x}\int str\left\{\hat{A_z} \,A_{\bz}\right\}
-\frac{\k}{2\pi x}\int bD[\hat{A_z}] c.
\ee
Assuming that the vectorial gauge symmetry is not anomalous (which may be
guaranteed by the existence of a nilpotent BRST charge) implies that
$W[\hat{A}_z]$,
\be
e^{\dis -W[\hat{A}_z]}=\int[dA_{\bz}][db][dc]e^{\dis -
\G[A_{\bz},\hat{A}_z,b,c]}
\ee
does not depend on the gauge, {\it i.e.} is independent of the value of
$\hat{A}_z$.
Performing the integral over the ghost fields using again the same
determinant as in eq.(\ref{yoyo}) yields
\be
\int[dA_{\bz}]e^{\dis -\G[A_{\bz}]+\frac{\k}{2\p x}\int \left\{ A_{\bz}\,
\hat{A_z}\right\}}=e^{\dis -(\k+2\tilde{h})S^-[g]},
\ee
where $\hat A_z=\del g g^{(-1)}$.
This corresponds to the value $Z_K=1/K$.

\setcounter{equation}{0}

\startappendix
\section{$\slt$ Embeddings}

Consider an embedding of $\slt$ in a (super) Lie algebra $\bar{g}$. The adjoint
representation of $\bar{g}$ branches into irreducible representations of
$\slt$. For a given embedding, we denote the generators of $\bar{g}$ by
$t_{(jm,\a_j)}$ where $j$, $2j\in{\bf N}$ labels the irreducible representation
of $\slt$, $m$ runs from $-j$ to $j$ and $\a_j$ counts the multiplicity of the
irreducible representation $j$ in the branching. The $\slt$ generators $e_\pm$
and $e_0$ are denoted by $e_\pm\equiv t_{(1\pm 1,0)}/\sqrt{2}$ and $e_0\equiv
t_{(10,0)}$. The $\slt$ algebra is given by $[e_0,e_\pm ]=\pm 2 e_\pm$ and
$[e_+,e_-]=e_0$. The action of the $\slt$ algebra on the other generators is
given by
\bea
[e_0,t_{(jm,\a_j)}]&=&2m\, t_{(jm,\a_j)}\nonu
[e_+, t_{(jm,\a_j)}]&=&(-)^{j+m}\sqrt{(j-m)(j+m+1)} t_{(jm+ 1,\a_j)}\nonu
[e_-, t_{(jm,\a_j)}]&=&(-)^{j+m-1}\sqrt{(j-m+1)(j+m)} t_{(jm- 1,\a_j)}
\eea
Computing $str ( [e_+, t_{(j\,-1/2)}][e_-, t_{(j\,1/2)}])$ in two different
ways, we conclude that, as the metric should be non-degenerate, {\it bosonic}
generators of half-integer spin always occur in pairs. This reflects the fact
that whenever bosonic multiplets of half-integer spin occur, the branching
contains a $U(1)$ generator under which the bosonic multiplets with $j$
halfinteger can be split into two subspaces having opposite eigenvalues under
the action of the $U(1)$ symmetry. We use the notation $t_{(jm,\a_j,\b_j)}$,
where $\b_j$ is the $U(1)$ eigenvalue.
The generators are normalized such that
\be
str \left( t_{(jm,\a_j,\b_j)}t_{(j'm',\a'_{j'},\b'_{j'})}\right)=
(-)^{2j(j-m)} (-)^{(\b_j)} 4 x y
\d_{jj'}\d_{m+m'}\d_{\a_j\a'_{j'}}\d_{\b_j+\b'_{j'}},\label{norm1}
\ee
where $(-)^{(\b_j)}=+1$ ($-1$) if $t_{(jm,\a_j,\b_j)}$ for $t$ bosonic and
$j$
halfinteger has positive (negative) chirality under the $U(1)$ and $y$ is the
index of embedding which is given by
\be
y=\frac{1}{3\tilde{h}}\sum_{j\a_j}(-)^{(\a_j)}j(j+1)(2j+1)\label{indexem}
\ee
and $(-)^{(\a_j)}=+1$ ($-1$) if $t_{(jm,\a_j)}$ is bosonic (fermionic).

In section six, we need the subalgebras\footnote{Throughout section 6, we will
use the symbol $\P$ for a projection operator acting on the Lie algebra
$\bar{g}$, the subindex indicates which part of the algebra survives.}
$\P_+\bar{g}$ and $\P_{\geq +1}\bar{g}$:
\bea
\P_+\bar{g}&=&\{ t_{(jm,\a)}|m > 0 ;\forall j,\a\}\nonu
\P_{\geq +1}\bar{g}&=&\{ t_{(jm,\a)}|m \geq 1 ;\forall j,\a\},
\eea
and we define $\P_{+1/2}\bar{g}$ as
\be
\P_{+1/2}\bar{g}=\{ t_{(j\frac 1 2 ,\a)}|\forall j,\a\}.
\ee
Analogous
definitions hold for $\P_-\bar{g}$, $\P_{\geq -1}\bar{g}$ and
$\P_{-1/2}\bar{g}$. Furthermore, we introduce upper indices $B$ and $F$ to
distinguish bosonic from fermionic generators, {\it e.g.}
$\P^B_{+1/2}\bar{g}=\{ t\in\P_{+1/2}\bar{g} |t\mbox{ bosonic}\}$ and
$\P^F_{+1/2}\bar{g}=\{ t\in\P_{+1/2}\bar{g} |t\mbox{ fermionic}\}$.
Finally $\P_{+1/2}\bar{g}$ is decomposed according to the $U(1)$ chirality as
$\P^{B}_{+1/2}\bar{g}=\P^{+B}_{+1/2}\bar{g}+\P^{-B}_{+1/2}\bar{g}$.



\frenchspacing
\begin{thebibliography}{11}
\bibitem{polyakov}
  A. M. Polyakov,  Mod. Phys. Lett. {\bf A2} (1987) 893
\bibitem{kpz}
  V.G. Knizhnik, A.M. Polyakov and A.B. Zamolodchikov,
  Mod.Phys.Lett. {\bf A3} (1988) 819
\bibitem{aleks}
  A. Alekseev and S. Shatashvili, \np {\bf B323} (1989) 719
\bibitem{bo}
  M. Bershadsky and H. Ooguri, \cmp {\bf 126} (1989) 49
\bibitem{ssvprep}
  K. Schoutens, A. Sevrin and P. van Nieuwenhuizen,
  in Proceedings of the Jan.\ 1991 Miami Workshop on {\it Quantum
  Field Theory, Statistical Mechanics, Quantum Groups and Topology},
  (Plenum, 1991)
\bibitem{brs2}
  K. Schoutens, A. Sevrin and P. van Nieuwenhuizen,
  Comm. Math. Phys. {\bf 124} (1989) 87
\bibitem{zamo}
  A.B. Zamolodchikov, Theor.Math.Phys. {\bf 63} (1985) 1205
\bibitem{pbks}
  P. Bouwknegt and K. Schoutens, preprint CERN-TH.6583/92 and ITP-SB-92-23, to
be published in Phys. Rep.
\bibitem{ssvna}
  K. Schoutens, A. Sevrin and P. van Nieuwenhuizen,
  Nucl. Phys. {\bf B364} (1991) 584
\bibitem{ssvnb}
  H. Ooguri, K. Schoutens, A. Sevrin and P. van Nieuwenhuizen,
  Comm. Math. Phys. {\bf 145} (1992) 515
\bibitem{ssvnc}
  K. Schoutens, A. Sevrin and P. van Nieuwenhuizen,
  Nucl. Phys. {\bf B371} (1992) 315
\bibitem{dbg}
  J. de Boer and J. Goeree, Utrecht preprint THU-92/33
\bibitem{grixu}
M. T. Grisaru and R. M. Xu, \pl {\bf 205B} (1988) 486.
\bibitem{sufrac}
 A. Polyakov and A. B. Zamolodchikov, Mod. Phys. Lett. {\bf A3} (1988) 1213.
\bibitem{bo2}
  M. Bershadsky and H. Ooguri, \pl {\bf 229B} (1989) 374.
\bibitem{osp1}
  V. G. Knizhnik, Theor. Math. Phys. {\bf 66} (1986) 68.
\bibitem{osp2}
  M. Bershadsky, \pl {\bf 174B} (1986) 285.
\bibitem{gustav}
   G.W. Delius, M.T. Grisaru and P. van Nieuwenhuizen, preprint
   CERN-TH.6458/92
\bibitem{n4}
 K. Schoutens, \np {\bf B295[FS21]} (1988) 634 \\
 A. Sevrin, W. Troost and A. Van Proeyen, \pl {\bf B208} (1988) 447
\bibitem{goddard} P. Goddard and A. Schwimmer, \pl {\bf 214B} (1988) 209
\bibitem{martin} M. Ro\v{c}ek and A. Sevrin, in preparation.
\bibitem{sorba} F. Delduc, E. Ragoucy and P. Sorba, Comm. Math. Phys. {\bf 146}
(1992) 403.
\bibitem{bv} J.A. Batalin and G.A. Vilkovisky, Phys. Rev. {\bf D28} (1983)
2567; {\bf D30} (1984) 508; Nucl. Phys. {\bf B234} (1984) 106.
\bibitem{proeyen} A. van Proeyen, preprint-KUL-TF-91/35, in the
proceedings of ``Strings and Symmetries 1991,'' World Scientific; W. Troost,
P. van Nieuwenhuizen and A. van Proeyen, Nucl. Phys. {\bf B333} (1990) 727;
W. Troost and A. Van Proeyen, {\it An introduction to
Batalin-Vilkovisky Lagrangian Quantisation},  Leuven
University Press, in preparation.
\bibitem{kz}
  V.G. Knizhnik and A.B. Zamolodchikov, \np {\bf B247} (1984) 83
\bibitem{ruud} A. Sevrin, R. Siebelink and W. Troost, in preparation.
\bibitem{kris1} A. Sevrin, K. Thielemans and W. Troost, preprint
LBL-33778, UCB-PTH-93/07, KUL-TF-93/10
\bibitem{zamo2}
  Al.B. Zamolodchikov, preprint ITEP 84-89 (1989)
\bibitem{PawMeiss} K.A. Meissner and J. Pawe\sss czyk, Mod. Phys. Lett. {\bf
A5} (1990) 763
\bibitem{bais} F. A. Bais, T. Tjin and P. van Driel, \np {\bf B357} (1991) 632.
\bibitem{dublin} L. Feher, L. O'Raifeartaigh, O. Ruelle, I. Tsutsui and
A. Wipf, DIAS -STP -91, UdeM -LPN -TH -71/91
\bibitem{alex} A. Deckmyn, \pl {\bf 298B} (1993) 318.
\bibitem{frenkel} B. L. Feigin and E. Frenkel, \pl {\bf 246B} (1990) 75.
\bibitem{tjin1} J. de Boer and T. Tjin, preprint THU-92-32.
\bibitem{tjin2} J. de Boer and T. Tjin, preprint THU-93-05.
\bibitem{bott} R. Bott and L. W. Tu, {\it Differential Forms in Algebraic
Topology}, Springer Verlag, 1986
\bibitem{bbss} F.A. Bais, P. Bouwknegt, M. Surridge and K. Schoutens, \np {\bf
B304} (1988) 348
\bibitem{spindel} A. Sevrin, W. Troost, A. Van Proeyen and P. Spindel, \np {\bf
B311} (1988) 465
\bibitem{nonrenor}M. T. Grisatu, W. Siegel and M. Ro\v{c}ek, \np {\bf B159}
(1979) 429; {\it Superspace}, S. J. Gates, M. T. Grisatu, W. Siegel and M.
Ro\v{c}ek, Benjamin/Cummings pub. comp. 1983, p358.
\bibitem{Sofra} L. Frappat, E. Ragoucy and P. Sorba, preprint
ENSLAPP-AL-391/92, july 1992.
\bibitem{josestany} J. M. Figueroa-O'Farrill and S. Schrans, \pl {\bf B257}
(1991) 69
\bibitem{alexalex} A. Deckmyn, A. Sevrin and W. Troost, work in progress
\bibitem{adem}
M. Ademollo et al., \pl {\bf 62B} (1976) 105; \np {\bf B111} (1976) 77; \np
{\bf B114} (1976) 297
\bibitem{golive} P. Goddard, D. Olive and G. Waterson, \cmp {\bf 112} (1987)
591.
\bibitem{dynkin} E. B. Dynkin, Amer. Math. Soc. Transl. {\bf 6} (1967) 111
\bibitem{mic} M. Bershadsky, \cmp {\bf 139} (1991) 71
\bibitem{wolf} M.~Berschadsky, W.~Lerche, D.~Nemeschansky and N.P.~Warner,
preprint CERN-TH.6694/92 and
W.~Lerche, preprint CERN-TH.6812/93
\bibitem{hv}
H. Verlinde,  \np {\bf B337} (1990) 652.
\bibitem{dbgcov} J. de Boer and J. Goeree, preprint THU-92-14.
\bibitem{saveliev} M.V. Savelev and A.M. Vershik Comm. Math. Phys. {\bf 126}
(1989) 367
\bibitem{polwi}
  A. M. Polyakov and P. B. Wiegmann, \pl {\bf 131B} (1983) 121;
  \pl {\bf 141B} (1984) 223.
\bibitem{wzw}
  E. Witten, \cmp {\bf 92} (1984) 455
\bibitem{orlando}
  O. Alvarez, \np {\bf B238} (1984) 61
\bibitem{dive}
  P. Di Vecchia and P. Rossi, \pl {\bf 140B} (1984) 344;
  P. Di Vecchia, B. Durhuus and J. L. Petersen, \pl {\bf 144B} (1984) 245.
\bibitem{poly2}
  A. M. Polyakov, Int. Jour. Mod. Phys. {\bf A5} (1990) 833
\bibitem{stony}
  K. Schoutens, A. Sevrin and P. van Nieuwenhuizen, in the proceedings of the
Stony Brook conference {\it Strings and Symmetries 1991}, (World Scientific,
1992).

\end{thebibliography}
\end{document}